\documentclass[10pt,twoside,ukrainian]{article}
\usepackage[T1]{fontenc}
\usepackage[utf8]{inputenc}
\usepackage[a4paper]{geometry}
\usepackage{amsmath}
\usepackage{amssymb}
\usepackage{setspace}

\makeatletter
\usepackage[english,ukrainian]{babel}
\usepackage{amsthm}
\usepackage[active]{srcltx}
\usepackage{cite}

\newtheorem{theorem}{Theorem}
\newtheorem{lemma}{Lemma}
\newtheorem{assertion}{Assertion}
\newtheorem{nasl}{Corollary}
\newtheorem{nasl*}{Наслідок *}
\newtheorem{properties}{Properties}
\newtheorem{definition}{Definition}
\theoremstyle{remark}
\newtheorem{rmk}{Remark}
\newtheorem{example}{{\bf\em Приклад}}

\newcommand{\BeginDef}{\begin{definition}}
\newcommand{\EndDef}  {\end{definition}}
\newcommand{\BeginLem}{\begin{lemma}}
\newcommand{\EndLem}  {\end{lemma}}
\newcommand{\BeginThm}{\begin{theorem}}
\newcommand{\EndThm}  {\end{theorem}}
\newcommand{\BeginAs}{\begin{assertion}}
\newcommand{\EndAs}{\end{assertion}}
\newcommand{\BeginNasl}{\begin{nasl}}
\newcommand{\EndNasl}  {\end{nasl}}
\newcommand{\BeginProp}{\begin{properties}}
\newcommand{\EndProp}{\end{properties}}
\newcommand{\BeginRmk}{\begin{rmk}}
\newcommand{\EndRmk}  {\end{rmk}}
\newcommand{\BeginProof}{\begin{proof}}
\newcommand{\EndProof}{\end{proof}}
\newcommand{\BeginEx}{\begin{example}}
\newcommand{\EndEx}{\end{example}}

\makeatother

\usepackage{babel}
\begin{document}

\author{Ya.I. Grushka}

\title{Abstract Coordinate Transforms in Kinematic Changeable Sets and their
Properties}

\date{\large  (Institute of Mathematics NAS of Ukraine, Kyiv, April, 2015) }

\maketitle
\large   \textbf{E-mail}: grushka@imath.kiev.ua

\textbf{}\bigskip{}
\bigskip{}


\setcounter{equation}{0}
\large 
\begin{center}
\parbox{15cm}{\normalsize \parindent=0.5cm 

\selectlanguage{english} 
\medskip 
\large 

One of the fundamental postulates of the special relativity theory is  existence of a single  system of universal coordinate transforms for inertial reference frames, that is coordinate transforms, which are uniquely determined by space-time coordinates of a material point. In this paper the abstract mathematical theory of coordinate transforms in kinematic changeable sets is developed. In particular it is proved the formal possibility of existence of kinematics, which do not allow universal coordinate transforms. Such kinematics may be applied for simulation the evolution of physical systems under the condition of hypothesis on existence of particle-dependent velocity of light. 

}
\end{center}

\medskip
\markboth{}{}
\sloppy
\allowdisplaybreaks \bigskip{}

\tableofcontents{}\newpage{}

\global\long\def\mathcircumflex{\mbox{\^{}}} 

\global\long\def\N{\mathbb{N}}
\global\long\def\Z{\mathbb{Z}}
\global\long\def\R{\mathbb{R}}
\global\long\def\CC{\mathbb{C}}
\global\long\def\card{\mathbf{card}}
\global\long\def\diag{\textsf{\textbf{diag}}}
\newcommand{\nlm}{\nolimits}\global\long\def\Dm{\mathfrak{D}}
\global\long\def\Rg{\mathfrak{R}}
\global\long\def\T{\mathbf{T}}
\global\long\def\TT{\mathbb{T}}
\global\long\def\MM{\mathcal{M}}
\global\long\def\arc#1{#1^{[-1]}}
\global\long\def\fff{\mathop{\leftarrow}}
\global\long\def\nff{\mathop{\:\not\not\negmedspace\negmedspace\fff}}
\global\long\def\AAA{\mathcal{A}}
\global\long\def\BBB{\mathcal{B}}
\global\long\def\Bs{\mathfrak{Bs}}
\global\long\def\Ind#1{\mathcal{I}nd\left(#1\right)}
\global\long\def\ind#1{\mathtt{ind}\left(#1\right)}
\global\long\def\Lk#1{\mathcal{L}k\left(#1\right)}
\global\long\def\cL{\mathcal{L}}
\global\long\def\cQ{\mathcal{Q}}
\global\long\def\cR{\mathcal{R}}
\global\long\def\Tm{\mathbf{Tm}}
\global\long\def\TM{\mathbb{T}\mathbf{m}}
\global\long\def\BS{\mathbb{B}\mathfrak{s}}
\global\long\def\bs#1{\mathsf{bs}\left(#1\right)}
\global\long\def\BsB{\Bs(\BBB)}
\global\long\def\BSB{\BS(\BBB)}
\global\long\def\TmB{\Tm(\BBB)}
\global\long\def\TMB{\TM(\BBB)}
\global\long\def\tm#1{\mathsf{tm}\left(#1\right)}
\global\long\def\BSBB#1{\BS\left(\BBB_{#1}\right)}
\global\long\def\bbL{\mathbb{L}}
\global\long\def\bbK{\mathbb{K}}
\global\long\def\At{\mathcal{A}t}
\global\long\def\fU{\mathfrak{U}}
\global\long\def\vcB{\overleftarrow{\BBB}}
\global\long\def\vfU{\overleftarrow{\fU}}
\global\long\def\cZ{\mathcal{Z}}
\global\long\def\Zim#1{\cZ\mathsf{im}\left(#1\right)}
\global\long\def\bbU{\mathbb{U}}
\global\long\def\IndZ{\Ind{\cZ}}
\global\long\def\LkZ{\Lk{\cZ}}
\global\long\def\un#1#2{\left\langle #2\fff#1\right\rangle }
\global\long\def\unn#1#2#3{\un{#1}{!\:#2}#3}
\global\long\def\ol#1{#1\mathcircumflex}
\global\long\def\oll{\ol{\lf}}
\global\long\def\w{\mathrm{w}}
\global\long\def\wx{\widetilde{\w}}
\global\long\def\fX{\mathfrak{X}}
\global\long\def\cX{\mathcal{X}}
\global\long\def\lsL{\bbL}
\global\long\def\spX{\fX}
\global\long\def\bX{\mathbf{X}}
\global\long\def\Lp#1#2{\mathfrak{Lp}\left(#1,#2\right)}
\global\long\def\kpQ{\mathfrak{Q}}
\global\long\def\tpT{\mathcal{T}}
\global\long\def\rfrm#1{\mathfrak{#1}}
\global\long\def\lf{\rfrm l}
\global\long\def\mf{\rfrm m}
\global\long\def\pf{\rfrm p}
\global\long\def\kf{\rfrm k}
\global\long\def\fP{\mathfrak{P}}
\global\long\def\fS{\mathfrak{S}}
\global\long\def\bbS{\mathbb{S}}
\global\long\def\Zimm#1{\cZ\mathsf{im}\left[#1\right]}
\global\long\def\Zk{\mathbf{Zk}}
\global\long\def\ZkQ{\Zk(\kpQ)}
\global\long\def\Tp{\mathcal{T}p}
\global\long\def\TpQ{\Tp(\kpQ)}
\global\long\def\Ls{\mathbb{L}s}
\global\long\def\LsQ{\Ls(\kpQ)}
\global\long\def\Ps{\mathfrak{Ps}}
\global\long\def\PsQ{\Ps(\kpQ)}
\global\long\def\ds{\mathbf{di}}
\global\long\def\dsQ{\ds_{\kpQ}}
\global\long\def\NQ#1{\left\Vert #1\right\Vert _{\kpQ}}
\global\long\def\ipQ#1#2{\left(#1,#2\right)_{\kpQ}}
\global\long\def\cG{\mathcal{G}}
\global\long\def\cK{\mathcal{K}}
\global\long\def\cG{\mathcal{G}}
\global\long\def\cK{\mathcal{K}}
\global\long\def\cC{\mathcal{C}}
\global\long\def\fC{\mathfrak{C}}
\global\long\def\bkK{\cK^{\mathfrak{b}}}
\global\long\def\bkC{\fC^{\mathfrak{b}}}
\global\long\def\cF{\mathcal{F}}
\global\long\def\cF{\mathcal{F}}
\global\long\def\obr#1#2{#1\upharpoonright#2}
\global\long\def\Cobr#1{\obr{\fC}{#1}}
\global\long\def\Col{\Cobr{\lf}}
\global\long\def\BE{\mathsf{BE}}
\global\long\def\BBE{\mathbb{BE}}
\global\long\def\BG{\mathsf{BG}}
\global\long\def\tr#1#2{\mathfrak{trj}_{#1}\left[#2\right]}
\global\long\def\TR{\mathbb{T}\mathbf{rj}}
\global\long\def\TRd{\overline{\TR}}
\global\long\def\kr{\mathfrak{q}}
\global\long\def\Mk{\mathbb{M}k}
\global\long\def\sPk{\mathbf{Q}}
\global\long\def\pk#1{\sPk^{\left\langle #1\right\rangle }}
\global\long\def\PKK#1#2{\sPk^{\un{#1}{#2}}}
\global\long\def\gteq{\mathop{\rightleftarrows}}
\global\long\def\ngteq{\mathop{\not\rightleftarrows}}
\global\long\def\I{\mathbb{I}}
\global\long\def\Kim#1{\mathfrak{Kim}\left(#1\right)}
\global\long\def\Kimm#1{\mathfrak{Kim}\left[#1\right]}
\global\long\def\LkC{\Lk{\fC}}
\global\long\def\bbUx{\widehat{\bbU}}
\global\long\def\Ha{\mathfrak{H}}
\global\long\def\MHa{\mathcal{M}\left(\Ha\right)}
\global\long\def\MccpH#1{\mathcal{M}_{#1,+}(\Ha)}
\global\long\def\McpHa{\MccpH c}
\global\long\def\oo{\mathbf{0}}
\global\long\def\e{\mathbf{e}_{0}}
\global\long\def\Tt{\widehat{\mathbf{T}}}
\global\long\def\Xx{\mathbf{X}}
\global\long\def\Ttt{\mathcal{T}}
\global\long\def\LHa{\cL\left(\Ha\right)}
\global\long\def\LHax{\cL^{\times}\left(\Ha\right)}
\global\long\def\LHao{\cL\left(\Ha_{1}\right)}
\global\long\def\LMHa{\cL\left(\MHa\right)}
\global\long\def\LMHax{\cL^{\times}\left(\MHa\right)}
\global\long\def\Hao#1{\Ha_{1}\left[#1\right]}
\global\long\def\Haoo#1{\Ha_{1\perp}\left[#1\right]}
\global\long\def\vl#1{\mathcal{V}\left(#1\right)}
\global\long\def\sn{\mathbf{span}\,}
\global\long\def\Xo#1{\Xx_{1}\left[#1\right]}
\global\long\def\Xoo#1{\Xx_{1}^{\perp}\left[#1\right]}
\global\long\def\UHa{\mathfrak{U}\left(\Ha_{1}\right)}
\global\long\def\BoHa{\mathbf{B}_{1}\left(\Ha_{1}\right)}
\global\long\def\Mkf{\mathsf{M}}
\global\long\def\Mkc#1{\Mkf_{c}\left(#1\right)}
\global\long\def\Mkk{\Mkf_{c}}
\global\long\def\bn{\mathbf{n}}
\global\long\def\ba{\mathbf{a}}
\global\long\def\Www#1#2{\mathbf{W}_{#1}\left[#2\right]}
\global\long\def\WsnJ#1{\Www{#1}{s,\bn,J}}
\global\long\def\WlsnJ{\WsnJ{\lambda,c}}
\global\long\def\Wwi#1{\Www{\infty,c}{#1}}
\global\long\def\WinJ{\Wwi{\bn,J}}
\global\long\def\WsnJa#1{\Www{#1}{s,\bn,J;\ba}}
\global\long\def\WlsnJa{\WsnJa{\lambda,c}}
\global\long\def\WLsnJa{\WsnJa{\lambda;\Vf}}
\global\long\def\Uuu#1#2{\mathbf{U}_{#1}\left[#2\right]}
\global\long\def\UsnJ#1{\Uuu{#1}{s,\bn,J}}
\global\long\def\UtsnJ{\UsnJ{\theta,c}}
\global\long\def\UsnJa#1{\Uuu{#1}{s,\bn,J,\ba}}
\global\long\def\UtsnJa{\UsnJa{\theta,c}}
\global\long\def\fio#1{\varphi_{0}\left(#1\right)}
\global\long\def\fiot{\fio{\theta}}
\global\long\def\fion#1{\varphi_{1}\left(#1\right)}
\global\long\def\fiont{\fion{\theta}}
\global\long\def\fioq#1{\varphi_{0}^{2}\left(#1\right)}
\global\long\def\fionq#1{\varphi_{1}^{2}\left(#1\right)}
\global\long\def\sign{\mathrm{sign}\,}
\global\long\def\LT#1{\mathfrak{OT}\left(#1\right)}
\global\long\def\LTp#1{\mathfrak{OT}_{+}\left(#1\right)}
\global\long\def\LTH{\LT{\Ha,c}}
\global\long\def\LTpH{\LTp{\Ha,c}}
\global\long\def\LG#1{\mathfrak{O}\left(#1\right)}
\global\long\def\LGp#1{\mathfrak{O}_{+}\left(#1\right)}
\global\long\def\LGH{\LG{\Ha,c}}
\global\long\def\LGpH{\LGp{\Ha,c}}
\global\long\def\LGm#1{\mathfrak{O}_{-}\left(#1\right)}
\global\long\def\PT#1{\mathfrak{PT}\left(#1\right)}
\global\long\def\PTp#1{\mathfrak{PT}_{+}\left(#1\right)}
\global\long\def\PTH{\PT{\Ha,c}}
\global\long\def\PTpH{\PTp{\Ha,c}}
\global\long\def\PTHC{\PT{\Ha;\Vf}}
\global\long\def\PTpHC{\PTp{\Ha;\Vf}}
\global\long\def\PG#1{\mathfrak{P}\left(#1\right)}
\global\long\def\PGp#1{\mathfrak{P}_{+}\left(#1\right)}
\global\long\def\PGm#1{\mathfrak{P}_{-}\left(#1\right)}
\global\long\def\PGH{\PG{\Ha,c}}
\global\long\def\PGpH{\PGp{\Ha,c}}
\global\long\def\Smxc#1#2{\sqrt{\left|1-\frac{#1^{2}}{#2^{2}}\right|}}
\global\long\def\Sxc#1#2{\sqrt{1-\frac{#1^{2}}{#2^{2}}}}
\global\long\def\smxc#1{\Smxc{#1}c}
\global\long\def\Vf{\mathfrak{V_{f}}}
\global\long\def\Cvp{\widetilde{\Vf}}
\global\long\def\MHac{\mathcal{M}\left(\Ha_{\Vf}\right)}
\global\long\def\Ipm#1#2{\I_{#1}\left[#2\right]}
\global\long\def\PkH{\mathbf{Pk}\left(\Ha\right)}
\global\long\def\Haa{\widehat{\Ha}}
\global\long\def\knK{\mathfrak{K}}
\global\long\def\KP#1{\knK\mathfrak{P}\left(#1\right)}
\global\long\def\KPn#1{\knK\mathfrak{P}_{0}\left(#1\right)}
\global\long\def\KPT#1{\knK\mathfrak{P}\mathfrak{T}\left(#1\right)}
\global\long\def\KPTn#1{\knK\mathfrak{P}\mathfrak{T}_{0}\left(#1\right)}
\global\long\def\xxv{\widetilde{x}}

\section{Introduction. }

\selectlanguage{english}

Due to the OPERA experiments conducted within 2011-2012 years \cite{OPERA01},
quite a lot physical works appeared, in which authors are trying to
modify the special relativity theory to agree its conclusions with
the hypothesis of existence of objects moving at velocity, greater
than the velocity of light. Despite the fact that the superluminal
results of OPERA experiments (2011-2012) were not confirmed later,
the problem of constructing a theory of super-light movement remains
actual within more than 50 last years \cite{Bilaniuk01,Bilaniuk02}.
At the present time existence of a few kinematic theories of tachyon
motion generates the problem of construction new mathematical structures,
which would allow to simulate of evolution of physical systems in
a framework of different laws of kinematics. Under the lack of experimental
verification of conclusions for tachyon kinematics theories, such
mathematical structures may at least guarantee the correctness of
receiving these conclusions in accordance with the postulates of these
theories. This paper is devoted to building of these mathematical
structures. Investigations in this direction may be also interesting
for astrophysics, because there exists the hypothesis, that in large
scale of the Universe, physical laws (in particular, the laws of kinematics)
may be different from the laws, acting in the neighborhood of our
solar System. 

On a physical level, the problem about investigation of kinematics
with arbitrary space-time coordinate transforms for inertial reference
frames, was presented in the \cite{BaccettiTateVisser} for the case,
when the space of geometric variables is three-dimensional and Euclidean.
The particular case of coordinate transforms, considered in \cite{BaccettiTateVisser}
are the (three-dimensional) classical Lorentz transforms as well as
generalized Lorentz transforms in the sense of E. Recami \cite{Recami1,Ricardo1,Hill_Cox}
(for reference frames moving at a velocity greater than the velocity
light). In the papers \cite{MyTmmTaxion01,MyTmmTaxion02} the general
definition of linear coordinate transforms and generalized Lorentz
transforms is given for the case, where the space of geometric variables
is any real Hilbert space. 

It should be noted, that mathematical apparatus of the papers \cite{BaccettiTateVisser,Hill_Cox,Ricardo1,Recami1,MyTmmTaxion01,MyTmmTaxion02}
is not based on the theory of changeable sets, which greatly reduces
its generality. In particular, mathematical apparatus of these papers
allows only studying of universal coordinate transforms (that is coordinate
transforms, which are uniquely determined by the geometrically-time
position of the considered object). The present paper is based on
the general theory of changeable sets \cite{MyTmm01,MyTmm03,MyTmm04,MyTmm06}.
In the paper the definitions of the actual and universal coordinate
transform in kinematic changeable sets are presented. We prove, that
in classical Galilean and Lorentz-Poincare kinematics the universal
coordinate transform always exists. Also in the paper we construct
the class of kinematics, in which every particle can have its own
``velocity of light'' and prove, that, in these kinematics, universal
coordinate transform does not exist.

\section{Basic Concepts of the Theory of Changeable Sets. }

In the present section we recall necessary denotations and concepts
of the changeable sets theory, introduced in \cite{MyTmm03} (see
also \cite{MyTmm01,MyTmm04,MyTmm05(YMJ),MyTmm06,MyTmm07,MyTmm08}).
Changeable and base changeable sets will be denoted by large calligraphy
or Gothic letters. Further for any base changeable set ~$\BBB$~
we use the following denotations:  \medskip{}

~~~ $\BsB$~-- the basic set or the set of all elementary states
of~ $\BBB$. 

~~~ $\BSB$~-- the set of all elementary-time states of ~$\BBB$. 

~~~ $\TmB$~-- the set of time points of~ $\BBB$. 

~~~ $\TMB=\left(\TmB,\leq_{\BBB}\right)$, where {\large{} }$\leq_{\BBB}$~
-- relation of time order of~ $\BBB$.{\large{} ~} 

~~~ $\fff\limits _{\BBB}$~-- directing relation of changes of~
$\BBB$. 

~~~ $\fff\limits _{\BBB}^{\BS}$~-- base of elementary processes
of~ $\BBB$.  \medskip{}

Elements of the sets $\BsB$ and $\BSB$ will be named by elementary
or elementary-time states of ~$\BBB$ (correspondingly).

\BeginProp[\cite{MyTmm08}] \label{Props:BMM(0)} Let $\BBB$ be
any base changeable set. Then: 
\begin{enumerate}
\item ~$\fff\limits _{\BBB}$ is reflexive binary relation, defined on
$\BsB$, that is for any elementary state $x\in\BsB$ the correlation
$x\fff\limits _{\BBB}x$ is performed. \label{enu:BMMProp(fff_Bs_reflexive)}
\label{enu:BMMProp(fff(Bs)_rel)}
\item $\leq_{\BBB}$ is relation of (not-strict) linear order defined on
$\TmB$ (i.e. $\TMB=\left(\TmB,\leq_{\BBB}\right)$ is linearly (totally)
ordered set in the sense of \emph{\cite[p. 12]{Birkhoff}}). 
\item $\BSB\subseteq\TmB\times\BsB$~ (where $\T\times\cX=\left\{ \left(t,x\right)\,|\: t\in\T,\, x\in\cX\right\} $
is the Cartesian product of the sets $\T$ and $\cX$). 
\item $\fff\limits _{\BBB}^{\BS}$ is reflexive binary relation, defined
on $\BSB$. \label{enu:BMMProp(fff(BS)_rel)} \label{enu:BMMProp(fff_reflexive)}
\end{enumerate}
\EndProp  

\BeginRmk \label{Rmk:BMM-Denot-Short}

In the case, when the base changeable set $\BBB$ is known in advance,
in the denotations $\leq_{\BBB}$, $\fff\limits _{\BBB}$, $\fff\limits _{\BBB}^{\BS}$
the char ``$\BBB$\,'' will be released, and we will use denotations
$\leq$, $\fff$, $\fff\limits ^{\BS}$ instead. In the cases, when
we can uniquely determine from the previous text, that the relation
$\fff\limits _{\BBB}^{\BS}$ is applied for elementary-time states
$\omega_{1},\omega_{2}\in\BSB$, we use the denotation $\omega_{2}\fff\limits _{\BBB}\omega_{1}$
or, simply, $\omega_{2}\fff\omega_{1}$ instead of $\omega_{2}\fff\limits _{\BBB}^{\BS}\omega_{1}$. 

\EndRmk \smallskip{}

Let, $\TT=\left(\T,\leq\right)$ be any linearly ordered set and $\cX$
be any set. For any ordered pair $\omega=\left(t,x\right)\in\T\times\cX$
we use the following denotations:
\begin{equation}
\bs{\omega}:=x,\quad\tm{\omega}:=t.\label{eq:bs()tm()def}
\end{equation}

\BeginProp[\cite{MyTmm03,MyTmm08}] \label{Props:BMM} Let $\BBB$,$\BBB_{1}$,$\BBB_{2}$
be any base changeable sets. Then: 
\begin{enumerate}
\item If $\omega_{1},\omega_{2}\in\BSB$ and $\omega_{2}\fff\omega_{1}$,
then \label{enu:BMMProp(fff,ffff)} $\bs{\omega_{2}}\fff\bs{\omega_{1}}$
and $\tm{\omega_{1}}\leq\tm{\omega_{2}}$. If, in addition, we have,
$\omega_{1}\neq\omega_{2}$, then $\tm{\omega_{1}}<\tm{\omega_{2}}$.
\label{enu:BMMProp(fff,bs,tm)} 
\item $\BsB=\left\{ \bs{\omega}\,|\,\omega\in\BSB\right\} $;\label{enu:BMMProp(Bs)}
\item For arbitrary $x_{1},x_{2}\in\BsB$ the condition $x_{2}\fff x_{1}$
holds if and only if there exist elementary-time states $\omega_{1},\omega_{2}\in\BSB$
such, that~  $\bs{\omega_{1}}=x_{1}$,~ $\bs{\omega_{2}}=x_{2}$
~and~ $\omega_{2}\fff\omega_{1}$. \label{enu:BMMProp(fff,Bs)}
\item If~ \textbf{$\TM\left(\BBB_{1}\right)=\TM\left(\BBB_{2}\right),\;\BSBB 1=\BSBB 2,\;\fff\limits _{\BBB_{1}}^{\BS}=\fff\limits _{\BBB_{2}}^{\BS}$},~
then~ $\BBB_{1}=\BBB_{2}$. \label{enu:aboutBMMequvalence}
\end{enumerate}
\EndProp  

\BeginRmk  

In some earlier works \cite{MyTmm03,MyTmmTaxion01} it was used the
synonymic term ``basic changeable set'' instead of the term ``base
changeable set''. 

\EndRmk  

Further for any changeable set ~$\cZ$~ we use the following denotations: 
\begin{description}
\item [{$\IndZ$~--}] the \textbf{\emph{index}} set of the changeable
set $\cZ$. {\large \vspace{-1mm}
}{\large \par}
\item [{$\LkZ$~--}] the set of all \textbf{\emph{areas of perception}}
of $\cZ$. 
\end{description}
Herewith for any areas of perception $\lf,\mf\in\LkZ$ we will use
denotations: 
\begin{description}
\item [{$\ind{\lf}$~--}] the index of the ~area of perception~~$\lf$.
 {\large \vspace{-1mm}
}{\large \par}
\item [{$\Bs(\lf)$,~$\BS(\lf)$,~$\Tm(\lf)$,~$\leq_{\lf}$,~$\fff\limits _{\lf}$,~$\fff\limits _{\lf}^{\BS}$~---}] the
set of all elementary states of $\lf$, the set of all elementary-time
states of $\lf$, the set of time points of $\lf$, relation of time
order of $\lf$, directing relation of changes of $\lf$, base of
elementary processes of $\lf$ (correspondingly). {\large \vspace{-1mm}
}{\large \par}
\item [{$\TM(\lf)=\left(\Tm(\lf),\leq_{\lf}\right)$.}] ~ {\large \vspace{-1mm}
}{\large \par}
\item [{$\un{\lf,\,\cZ}{\mf}$~--}] \textbf{\emph{unification mapping~}}
from the area of perception ~$\lf$~ into the area of perception
~$\mf$. 
\end{description}
\BeginRmk  \label{Rmk:BChSet_to_lf} 

From the definition of changeable set (\cite[Definition 3.1]{MyTmm06},
\cite[Definition 9.6]{MyTmm03}) it directly follows, that for any
area of perception $\lf\in\LkZ$ of any changeable set $\cZ$,  Properties
\ref{Props:BMM(0)} and Properties \ref{Props:BMM} are holding, \label{enu:BChSet_lf_Properties}
where \emph{we use all abbreviated variants of notations, described
in the Remark \ref{Rmk:BMM-Denot-Short} }(but, with replacement of
the symbol ``$\BBB$\,'' by the symbol ``$\lf$'' and the term ``base
changeable set'' by the term ``area of perception''). 

\EndRmk  

\BeginProp[\cite{MyTmm08}] \label{Props:ChSets(0)}

Let $\cZ$ be any changeable set. On that: 
\begin{enumerate}
\item The sets $\LkZ$ and $\IndZ$ always are nonempty, moreover $\IndZ=\left\{ \ind{\lf}\,|\:\lf\in\LkZ\right\} $.
\label{enu:LkZnoempty}
\item For arbitrary~ $\lf,\mf\in\LkZ$~ the equality~  $\lf=\mf$~ holds
if and only if ~$\ind{\lf}=\ind{\mf}$. 
\item For arbitrary $\lf,\mf\in\LkZ$ the unification mapping $\un{\lf,\,\cZ}{\mf}$
is the mapping from $2^{\BS(\lf)}$ into $2^{\BS(\mf)}$, where $2^{\mathbf{M}}$
means the set of all subset of the set $\mathbf{M}$. 
\end{enumerate}
\EndProp 

\BeginAs[\cite{MyTmm08}] \label{As:Z1=00003DZ2_(Lk,unif)} 

Let, $\cZ_{1}$, $\cZ_{2}$ be arbitrary changeable sets, moreover,
$\Lk{\cZ_{1}}=\Lk{\cZ_{2}}$ and for any areas of perception $\lf,\mf\in\Lk{\cZ_{1}}=\Lk{\cZ_{2}}$
it is true the equality: $\un{\lf,\cZ_{1}}{\mf}=\un{\lf,\cZ_{2}}{\mf}$.
Then, $\cZ_{1}=\cZ_{2}$. 

\EndAs 

Let $\lf,\mf\in\LkZ$ be arbitrary areas of perception of the changeable
set $\cZ$. Then for any subset $A\subseteq\BS(\lf)$ we denote by
$\un{\lf,\,\cZ}{\mf}A$~ the result of action of unification mapping
$\un{\lf,\,\cZ}{\mf}$ on the set $A$. Hence, $\un{\lf,\,\cZ}{\mf}A:=\un{\lf,\,\cZ}{\mf}(A)$.
In the case, when the changeable set $\cZ$ is known in advance, we
will release the symbol ``$\cZ$'' in the denotation $\un{\lf,\,\cZ}{\mf}$,
using instead of it the denotation $\un{\lf}{\mf}$. 

In the papers \cite{MyTmm03,MyTmm06} it is proved, that for any changeable
set $\cZ$ the following properties are performed: 

\BeginProp  \label{Props:ChSets(1)}
\begin{enumerate}
\item Any area of perception $\lf\in\LkZ$ can be represented in the form
of ordered pair,~ $\lf=\left(\ind{\lf},\oll\right)$,~ where ~$\oll$~
is the base changeable set. Moreover:\\
 $\Bs(\lf)=\Bs\left(\oll\right)$, $\BS(\lf)=\BS\left(\oll\right)$,
$\Tm(\lf)=\Tm\left(\oll\right)$, $\TM(\lf)=\TM\left(\oll\right)$,
\\
$\leq_{\lf}=\leq_{\oll}$, $\fff\limits _{\lf}=\fff\limits _{\oll}$,
$\fff\limits _{\lf}^{\BS}=\fff\limits _{\oll}^{\BS}$. \label{enu:PropChSets:Oll}
\label{enu:PropChSetsBegin} {\large \vspace{-1mm}
}{\large \par}
\item For arbitrary $\lf\in\LkZ$ and $A\subseteq\BS(\lf)$ it is true the
equality, $\un{\lf}{\lf}A=A$. \label{enu:PropChSets:<ll>unif}{\large \vspace{-1mm}
}{\large \par}
\item If  $\lf,\mf\in\LkZ$ and $A\subseteq B\;\subseteq\BS(\lf)$, then
$\un{\lf}{\mf}A\subseteq\un{\lf}{\mf}B$; {\large \vspace{-1mm}
}{\large \par}
\item For arbitrary $\lf,\mf,\pf\in\LkZ$ and  $A\subseteq\BS(\lf)$ it
holds the inclusion: \label{enu:PropChSetsEnd} 
\[
\un{\mf}{\pf}\un{\lf}{\mf}A\subseteq\un{\lf}{\pf}A.
\]

\end{enumerate}
\EndProp  

\BeginDef[\cite{MyTmm08}] \label{Def:fvi} 

Changeable set $\cZ$ is named \textbf{precisely visible} if and only
if for arbitrary $\lf,\mf\in\LkZ$ and $A\subseteq\BS(\lf)$ the condition
$A\neq\emptyset$ leads to $\un{\lf}{\mf}A\neq\emptyset$. 

\EndDef

\BeginAs[\cite{MyTmm04,MyTmm03}] Let $\lf,\mf\in\LkZ$ be arbitrary
areas of perception of any precisely visible changeable set $\cZ$.
Then for any $\omega\in\BS(\lf)$ the unique element $\omega'=:\unn{\lf,\cZ}{\mf}{\omega}\in\BS(\mf)$
exists such, that $\left\{ \omega'\right\} =\un{\lf}{\mf}\left\{ \omega\right\} $. 

\EndAs 

Thus, in any precisely visible changeable set $\cZ$ for arbitrary
$\lf,\mf\in\LkZ$ and $\omega\in\BS(\lf)$ it is performed the equality:
\[
\left\{ \unn{\lf,\cZ}{\mf}{\omega}\right\} =\un{\lf}{\mf}\left\{ \omega\right\} .
\]
 The mapping $\unn{\lf,\cZ}{\mf}:\BS(\lf)\mapsto\BS(\mf)$ is referred
to as \textbf{\emph{precise unification}} mapping. In the cases, when
the changeable set $\cZ$ is known in advance, we use the denotation
$\unn{\lf}{\mf}{\,}$ instead of the denotation $\unn{\lf,\cZ}{\mf}{\,}$. 

\BeginAs[\cite{MyTmm04,MyTmm03}]\label{As:FviUniBijection}

Let $\lf,\mf\in\LkZ$, where $\cZ$ is precisely visible changeable
set. Then the sets $\BS(\lf)$ and~ $\BS(\mf)$ are equipotent. Moreover,
the mapping: $f(\omega)=\unn{\lf}{\mf}{\omega}$ ($\omega\in\BS(\lf)$)
is bijection (one-to-one correspondence) between these sets. 

\EndAs 

Using Property \ref{Props:ChSets(1)}(\ref{enu:PropChSets:<ll>unif}),
as well as \cite[theorems 5.2,~5.1]{MyTmm04} or \cite[theorems 11.2, 11.1]{MyTmm03},
we receive the following properties of precise unification mappings
in precisely visible changeable sets. 

\BeginProp  \label{Props:FviUni} 

Let $\cZ$ be any precisely visible changeable set, and $\lf,\mf,\pf\in\LkZ$
be arbitrary areas of perception of $\cZ$. Then: 
\begin{enumerate}
\item $\forall\:\omega\in\BS(\lf)$~ $\unn{\lf}{\lf}{\omega}=\omega$;\label{enu:FviUni<l,l>}
\item $\forall\: A\subseteq\BS(\lf)$~ $\un{\lf}{\mf}A=\left\{ \unn{\lf}{\mf}{\omega}\,|\:\omega\in A\right\} $;$\qquad$

\item $\forall\:\omega\in\BS(\lf)$~ $\unn{\mf}{\pf}{\unn{\lf}{\mf}{\omega}}=\unn{\lf}{\pf}{\omega}$.$\qquad$\label{enu:FviUniTranstive} 
\end{enumerate}
\EndProp

\section{Changeable Sets and Kinematics. }

\subsection{Mathematical Objects for Constructing of Geometric Environments of
Changeable Sets.}

This subsection is purely technical in nature. In this subsection
we don't introduce any essentially new notions. But we try to include
the most frequently used mathematical spaces, which at least somehow
related to geometry, into single mathematical structure, which will
be convenient for further construction of abstract kinematics. 

\BeginDef 

The ordered triple $\lsL=\left(\bbK,\oplus,\otimes\right)$ will be
named the \textbf{linear structure} over the set $\spX$, if and only
if: 
\begin{enumerate}
\item $\bbK=\left(\mathbf{K},+_{\bbK},\times_{\bbK}\right)$ is a field.
{\large \vspace{-1mm}
}{\large \par}
\item $\oplus:\spX\times\spX\mapsto\spX$ is a binary operation over $\spX$;
{\large \vspace{-1mm}
}{\large \par}
\item $\otimes:\mathbf{K}\times\spX\mapsto\spX$ is a binary operation,
acting from $\mathbf{K}\times\spX$ into $\spX$. {\large \vspace{-1mm}
}{\large \par}
\item The ordered triple $\left(\spX,\oplus,\otimes\right)$ is a linear
space over the field $\bbK$. 
\end{enumerate}
In the case, when $\bbK\in\left\{ \R,\CC\right\} $, the linear structure
$\lsL$ will be named as \textbf{numerical linear structure} over
$\spX$. 

\EndDef 

Let $\lsL=\left(\bbK,\oplus,\otimes\right)$ be a linear structure
over $\spX$. In this case the linear space over the field $\bbK$,
generated by $\lsL$ will be denoted by $\Lp{\spX}{\lsL}$ ~ ($\Lp{\spX}{\lsL}=\left(\spX,\oplus,\otimes\right)$). 

The next definition is based on the conception, that the majority
of the most frequently used mathematical objects (including functions,
relations, algebraic operations, ordered pairs or compositions) are
sets. 

\BeginDef 

An ordered composition of six sets $\kpQ=\left(\spX,\tpT,\lsL,\rho,\left\Vert \cdot\right\Vert ,\left(\cdot,\cdot\right)\right)$
will be named as \textbf{coordinate space}, if and only if the following
conditions are satisfied: 
\begin{enumerate}
\item $\spX\neq\emptyset$. {\large \vspace{-1mm}
}{\large \par}
\item $\tpT\cup\lsL\neq\emptyset$. {\large \vspace{-1mm}
}{\large \par}
\item If $\tpT\neq\emptyset$, then $\tpT$ is topology over $\spX$.{\large \vspace{-1mm}
}{\large \par}
\item If~ $\lsL\neq\emptyset$, then $\lsL$ is numerical linear structure
over $\spX$. {\large \vspace{-1mm}
}{\large \par}
\item If $\lsL\neq\emptyset$ and $\tpT\neq\emptyset$, then the pair $\left(\Lp{\spX}{\lsL},\tpT\right)$
is a linear topological space. {\large \vspace{-1mm}
}{\large \par}
\item If $\rho\neq\emptyset$, then:\label{enu:Metr}{\large \vspace{-0.5mm}
}{\large \par}

\begin{description}
\item [{\ref{enu:Metr}.1)}] $\rho$ is the metrics over $\spX$;
\item [{\ref{enu:Metr}.2)}] $\tpT\neq\emptyset$ ~and~ the topology
$\tpT$ is generated by the metrics $\rho$. {\large \vspace{-1mm}
}{\large \par}
\end{description}
\item If~ $\left\Vert \cdot\right\Vert \neq\emptyset$, then: \label{enu:Norm}{\large \vspace{-1mm}
}{\large \par}

\begin{description}
\item [{\ref{enu:Norm}.1)}] $\lsL\neq\emptyset$~ and~ $\left\Vert \cdot\right\Vert $
is the norm on the linear space $\Lp{\spX}{\lsL}$; 
\item [{\ref{enu:Norm}.2)}] $\rho\neq\emptyset$~ and~ the metrics $\rho$
is generated by the norm $\left\Vert \cdot\right\Vert $. {\large \vspace{-1mm}
}{\large \par}
\end{description}
\item If $\left(\cdot,\cdot\right)\neq\emptyset$, then: \label{enu:Iprod}{\large \vspace{-1mm}
}{\large \par}

\begin{description}
\item [{\ref{enu:Iprod}.1)}] $\left\Vert \cdot\right\Vert \neq\emptyset$
(and hence, according to \textbf{\emph{\ref{enu:Norm}.1)}}, $\lsL\neq\emptyset$); 
\item [{\ref{enu:Iprod}.2)}] $\left(\cdot,\cdot\right)$ is the inner
product on the linear space $\Lp{\spX}{\lsL}$; 
\item [{\ref{enu:Iprod}.3)}] the norm $\left\Vert \cdot\right\Vert $
is generated by the inner product $\left(\cdot,\cdot\right)$. {\large \vspace{-1mm}
}{\large \par}
\end{description}
\end{enumerate}
\EndDef

\subparagraph{Notes on denotations. }

Let $\kpQ=\left(\spX,\tpT,\lsL,\rho,\left\Vert \cdot\right\Vert ,\left(\cdot,\cdot\right)\right)$
be a coordinate space, where in the case $\lsL\neq\emptyset$ we have,
that $\lsL=\left(\bbK,\oplus,\otimes\right)$ is a numerical linear
structure over $\spX$. Further we will use the following denotations: 
\begin{enumerate}
\item $\ZkQ:=\spX$\quad{} (the set $\ZkQ$ will be named the \textbf{\emph{set
of coordinate values}} of $\kpQ$). 
\item $\TpQ:=\tpT$ \quad{} ($\TpQ$ will be referred to as \textbf{\emph{topology}}
of $\kpQ$). 
\item $\LsQ:=\lsL$\quad{} ($\LsQ$ will be named the \textbf{\emph{linear
structure }}of $\kpQ$). 
\item $\PsQ:=\begin{cases}
\bbK, & \LsQ\neq\emptyset\\
\emptyset, & \LsQ=\emptyset
\end{cases}$\quad{} ($\PsQ$ will be referred to as  \textbf{\emph{field of scalars}}
of $\kpQ$).
\item For the elements $x_{1},\dots,x_{n}\in\ZkQ$, $\lambda_{1},\dots,\lambda_{n}\in\PsQ$
($n\in\N)$ we will use the denotation, $\left(\lambda_{1}x_{1}+\cdots+\lambda_{n}x_{n}\right)_{\kpQ}:=\lambda_{1}\otimes x_{1}\oplus\cdots\oplus\lambda_{n}\otimes x_{n}$. 
\item $\dsQ:=\rho$\quad{} ($\dsQ$ will be named the \textbf{\emph{distance}}
on  $\kpQ$). 
\item $\NQ{\cdot}:=\left\Vert \cdot\right\Vert $\quad{} ($\NQ{\cdot}$
will be named the \textbf{\emph{norm}} on $\kpQ$). 
\item $\ipQ{\cdot}{\cdot}:=\left(\cdot,\cdot\right)$\quad{} ($\ipQ{\cdot}{\cdot}$
will be referred to as \textbf{\emph{inner product}} on $\kpQ$). 
\end{enumerate}
Elements of kind $x\in\ZkQ$ will be named as \textbf{\emph{coordinates}}
of the coordinate space $\kpQ$, also, in the case $\LsQ\neq\emptyset$
we will name these elements as \textbf{\emph{vectors (vector coordinates)}}
of $\kpQ$. Where it does not cause confusion the symbol ``$\kpQ$''
in the denotations $\left(\lambda_{1}x_{1}+\cdots+\lambda_{n}x_{n}\right)_{\kpQ}$,
$\dsQ$, $\NQ{\cdot}$, $\ipQ{\cdot}{\cdot}$ will be released, and
we will use the abbreviated denotations $\lambda_{1}x_{1}+\cdots+\lambda_{n}x_{n}$,
$\ds$, $\left\Vert \cdot\right\Vert $, $\left(\cdot,\cdot\right)$
correspondingly.

\subsection{Kinematic Changeable Sets. \label{sub:KinChSets} }

\BeginDef  \label{Def:KinSets}
\begin{enumerate}
\item The pair $\cG_{0}=\left(\kpQ,k\right)$ we name \textbf{geometric
environment} of base changeable set\textbf{ }$\BBB$, if and only
if:

\begin{description}
\item [{a)}] $\kpQ$ is a coordinate space;
\item [{b)}] $k:\,\BsB\mapsto\ZkQ$ is a mapping from $\BsB$ into $\ZkQ$. 
\end{description}

In this case the pair $\bkC=\left(\BBB,\cG_{0}\right)=\left(\BBB,\left(\kpQ,k\right)\right)$
we name by \textbf{base kinematic changeable set}, or, abbreviated,
by \textbf{base kinematic }set. 

\item Let $\cZ$ be any changeable set. An indexed family of pairs~ $\cG=\left(\left(\kpQ_{\lf},k_{\lf}\right)\,|\:\lf\in\LkZ\right)$
will be named \textbf{geometric environment} of the changeable set
$\cZ$, if and only if for any area of perception $\lf\in\LkZ$ the
ordered pair $\left(\kpQ_{\lf},k_{\lf}\right)$ is geometric environment
of the base changeable set~ $\oll$, generated by the area of perception
$\lf$, i.e. if and only if the pair $\left(\oll,\left(\kpQ_{\lf},k_{\lf}\right)\right)$
is a base kinematic changeable set for an arbitrary $\lf\in\LkZ$.
\medskip{}
\\
In this case we name the pair $\fC=\left(\cZ,\cG\right)$ by \textbf{kinematic
changeable set}, or, abbreviated, by \textbf{kinematic} \textbf{set}.
\label{enu:Def:KSets:General}\vspace{-1mm}

\end{enumerate}
\EndDef 

Note, that in this article we consider only kinematic sets with constant
(unchanging over time) geometry. These kinematic sets are sufficient
for construction of abstract kinematics in inertial reference frames.
If we make a some modification of the Definition \ref{Def:KinSets},
we will be able to define also kinematic sets with variable (over
time) geometry (i.e., in principle, this is, possible to do).

\subsubsection{System of Denotations for Base Kinematic Sets. }

Let, $\bkC=\left(\BBB,\cG_{0}\right)$ be any base kinematic set (where
$\cG_{0}=\left(\kpQ,k\right)$). Henceforth we use the following system
of denotations. 

a) Denotations, induced from the theory of base changeable sets: 
\begin{gather*}
\begin{array}{lll}
\Bs\left(\bkC\right):=\BsB; & \BS\left(\bkC\right):=\BSB;\\
\fff\limits _{\bkC}:=\fff\limits _{\BBB}; & \fff\limits _{\bkC}^{\BS}:=\fff\limits _{\BBB}^{\BS};\\
\Tm\left(\bkC\right):=\TmB; & \TM\left(\bkC\right):=\TMB; & \leq_{\bkC}\;:=\;\leq_{\BBB}.
\end{array}
\end{gather*}

b) Denotations, induced from the denotations for coordinate spaces:
\begin{gather*}
\begin{array}{lll}
\Zk\left(\bkC\right):=\Zk\left(\kpQ\right); & \Tp\left(\bkC\right):=\Tp\left(\kpQ\right); & \Ls\left(\bkC\right):=\Ls\left(\kpQ\right);\\
\Ps\left(\bkC\right):=\Ps\left(\kpQ\right); & \ds_{\bkC}:=\ds_{\kpQ}; & \left\Vert \cdot\right\Vert _{\bkC}:=\left\Vert \cdot\right\Vert _{\kpQ};\\
\left(\cdot,\cdot\right){}_{\bkC}:=\left(\cdot,\cdot\right)_{\kpQ}.
\end{array}
\end{gather*}

\noindent Also in the case~ $\Ls\left(\bkC\right)\neq\emptyset$~
for arbitrary $a_{1},\dots,a_{n}\in\Zk\left(\bkC\right)$, $\lambda_{1},\dots\lambda_{n}\in\Ps\left(\bkC\right)$
we use the denotation, $\left(\lambda_{1}a_{1}+\cdots+\lambda_{n}a_{n}\right)_{\bkC}:=\left(\lambda_{1}a_{1}+\cdots+\lambda_{n}a_{n}\right)_{\cQ}$. 

c) Own designations for base kinematic sets:
\begin{gather*}
\BE\left(\bkC\right):=\BBB;\qquad\BG\left(\bkC\right):=\kpQ;\qquad\kr_{\bkC}(x):=k(x)\;\left(x\in\Bs\left(\bkC\right)\right).
\end{gather*}
Note, that for any elementary state $x\in\Bs\left(\bkC\right)$ the
function $\kr_{\bkC}(\cdot)$ puts in accordance its coordinate $\kr_{\bkC}(x)\in\Zk\left(\bkC\right)$.
    \medskip{}

d) Abbreviated version of denotations \vspace{-2mm}

\begin{itemize}
\item We use all abbreviated variants of denotations, described in the
Remark \ref{Rmk:BMM-Denot-Short} (but, with the replacement of the
symbol ``$\BBB$\,'' by the symbol ``$\bkC$'' and the term ``base
changeable set'' by the term ``base kinematic set''. \vspace{-1.5mm}

\item In the cases, when the base kinematic set $\bkC$ is known in advance,
we will use the denotations $\ds$, $\left\Vert \cdot\right\Vert $,
$\left(\cdot,\cdot\right)$, $\kr(x)$ instead of the denotations~
$\ds_{\bkC}$, $\left\Vert \cdot\right\Vert _{\bkC}$, $\left(\cdot,\cdot\right){}_{\bkC}$,
$\kr_{\bkC}(x)$ (correspondingly).
\end{itemize}

\subsubsection{System of Denotations for Kinematic Sets. \label{sub:fKMMDenotations}}

Let, $\fC=\left(\cZ,\cG\right)$, where $\cG=\left(\left(\kpQ_{\lf},k_{\lf}\right)\,|\:\lf\in\LkZ\right)$
be any kinematic set. 
\begin{description}
\item [{a)}] The changeable set~ $\BBE\left(\fC\right):=\cZ$~ will be
named the \textbf{\emph{evolution base}} of the kinematic set $\fC$.
 
\item [{b)}] The sets~ $\Ind{\fC}:=\IndZ=\Ind{\BBE\left(\fC\right)};$~
$\LkC:=\LkZ=\Lk{\BBE\left(\fC\right)}$~ will be named the set of
\textbf{\emph{indexes}} and the the set of all \textbf{\emph{reference
frames}} of kinematic set $\fC$ (correspondingly). Note, that in
the case of kinematic sets (in contrast to the case of changeable
sets) we use the term ``reference frame'' instead of the term ``area
of perception''. 
\item [{c)}] For any reference frame  $\lf\in\LkC=\LkZ$ we keep all denotations,
introduced for area of perception of changeable sets (it concerns
the denotations: $\ind{\lf}$, $\Bs(\lf)$, $\fff\limits _{\lf}$,
$\BS(\lf)$, $\fff\limits _{\lf}^{\BS}$, $\Tm(\lf)$, $\TM(\lf)$,
$\leq_{\lf}$).  
\item [{d)}] For arbitrary reference frames  $\lf,\mf\in\LkC$ it is induced
the denotation for unification mapping: 
\[
\un{\lf,\,\fC}{\mf}:=\un{\lf,\,\cZ}{\mf}.
\]
 In particular in the case, when the changeable set $\cZ$ is precisely
visible (in this case we say, that the kinematic set $\fC$ is \textbf{\emph{precisely
visible}}), we introduce the denotation:
\[
\unn{\lf,\fC}{\mf}{\,}:=\unn{\lf,\cZ}{\mf}.
\]

\item [{e)}] For any reference frame $\lf\in\LkC$ we introduce the denotation
$\Col=\left(\oll,\left(\kpQ_{\lf},k_{\lf}\right)\right)$. By the
Definition \ref{Def:KinSets}, the pair $\Col$ is a base kinematic
set (for arbitrary reference frame $\lf\in\LkC$). The base kinematic
set $\Col$ will be named the \textbf{\emph{image of kinematic set
~$\fC$~}} in the reference frame~$\lf$. 
\item [{f)}] For any reference frame  $\lf\in\LkC$ we introduce the following
denotations:
\[
\begin{array}{ll}
\Zk(\lf;\,\fC):=\Zk\left(\Col\right)=\Zk\left(\kpQ_{\lf}\right); & \Ls(\lf;\,\fC):=\Ls\left(\Col\right)=\Ls\left(\kpQ_{\lf}\right);\\
\Tp(\lf;\,\fC):=\Tp\left(\Col\right)=\Tp\left(\kpQ_{\lf}\right); & \Ps(\lf;\,\fC):=\Ps\left(\Col\right)=\Ps\left(\kpQ_{\lf}\right);\\
\left\Vert \cdot\right\Vert _{\lf,\fC}:=\left\Vert \cdot\right\Vert _{\Col}=\left\Vert \cdot\right\Vert _{\kpQ_{\lf}}; & \ds_{\lf}\left(\cdot;\,\fC\right):=\ds_{\Col}=\ds_{\kpQ_{\lf}};\\
\left(\cdot,\cdot\right)_{\lf,\fC}:=\left(\cdot,\cdot\right)_{\Col}=\left(\cdot,\cdot\right)_{\kpQ_{\lf}}; & \BE(\lf):=\BE(\Col)=\oll;\\
 & \BG(\lf;\,\fC):=\BG(\Col)=\kpQ_{\lf}.
\end{array}
\]
Also for reference frames $\lf\in\LkC$ such, that $\Ls(\lf)\neq\emptyset$
and for arbitrary $a_{1},\dots,a_{n}\in\Zk(\lf;\,\fC)$, $\lambda_{1},\dots\lambda_{n}\in\Ps(\lf;\,\fC)$
we will use the denotation, $\left(\lambda_{1}a_{1}+\cdots+\lambda_{n}a_{n}\right)_{\lf,\fC}:=\left(\lambda_{1}a_{1}+\cdots+\lambda_{n}a_{n}\right)_{\cQ_{l}}$. 
\item [{g)}] For any reference frame $\lf\in\LkC$ we use the following
denotation: 
\begin{align*}
\kr_{\lf}\left(x;\,\fC\right) & :=\kr_{\Col}\left(x\right)=k_{\lf}(x),\; x\in\Bs\left(\lf\right).
\end{align*}

\item [{h)}] Abbreviated versions of denotations:  \end{description}
\begin{itemize}
\item In the cases, when the kinematic set $\fC$ is known in advance, we
will use the denotations $\un{\lf}{\mf}$, $\unn{\lf}{\mf}{\,}$,
$\Zk(\lf)$, $\Ls(\lf)$, $\ds_{\lf}$, $\left(\cdot,\cdot\right)_{\lf}$,
$\Tp(\lf)$, $\Ps(\lf)$, $\left\Vert \cdot\right\Vert _{\lf}$, $\BG(\lf)$,
$\kr_{l}\left(x\right)$ instead of the denotations~$\un{\lf,\,\fC}{\mf}$,
$\unn{\lf,\,\fC}{\mf}{\,}$, $\Zk(\lf;\,\fC)$, $\Ls(\lf;\,\fC)$,
$\ds_{\lf}\left(\cdot;\,\fC\right)$, $\left(\cdot,\cdot\right)_{\lf,\fC}$,
$\Tp(\lf;\,\fC)$, $\Ps(\lf;\,\fC)$, $\left\Vert \cdot\right\Vert _{\lf,\fC}$,
$\BG(\lf;\,\fC)$, $\kr_{l}\left(x;\,\fC\right)$ (correspondingly). 
\item In the cases, when the reference frame $\lf\in\LkC$ is known in advance,
we will use the denotations $\ds$, $\left\Vert \cdot\right\Vert $,
$\left(\cdot,\cdot\right)$, $\kr(x)$, $\lambda_{1}a_{1}+\cdots+\lambda_{n}a_{n}$
instead of the denotations~ $\ds_{\lf}$, $\left\Vert \cdot\right\Vert _{\lf}$,
$\left(\cdot,\cdot\right){}_{\lf}$, $\kr_{\lf}\left(x\right)$, $\left(\lambda_{1}a_{1}+\cdots+\lambda_{n}a_{n}\right)_{\lf,\fC}$
(correspondingly). Also we use all abbreviated variants of denotations
for areas of perception of changeable sets, described in the Remark
\ref{Rmk:BChSet_to_lf}. 
\end{itemize}

\BeginAs[\cite{MyTmm08}]\label{As:Nasl:C1=00003DC2_(Lk,BG(l),ql,unif)}

Let $\fC_{1}$, $\fC_{2}$ be arbitrary kinematic sets, and besides: 
\begin{enumerate}
\item $\Lk{\fC_{1}}=\Lk{\fC_{2}}$. 
\item For any reference frame $\lf\in\Lk{\fC_{1}}=\Lk{\fC_{2}}$ they are
true the equalities, $\BG\left(\lf;\,\fC_{1}\right)=\BG\left(\lf;\,\fC_{2}\right)$
~and~ $\kr_{\lf}\left(x,\,\fC_{1}\right)=\kr_{\lf}\left(x,\,\fC_{2}\right)\;\left(\forall x\in\Bs(\lf)\right)$. 
\item For arbitrary reference frames $\lf,\mf\in\Lk{\fC_{1}}=\Lk{\fC_{2}}$
it is true the equality, $\un{\lf,\fC_{1}}{\mf}=\un{\lf,\fC_{2}}{\mf}$. 
\end{enumerate}
Then, $\fC_{1}=\fC_{2}$. 

\EndAs

\BeginRmk 

From the system of denotations, accepted in the subsection \ref{sub:KinChSets},
it follows, that for any kinematic set $\fC$, Properties \ref{Props:ChSets(0)},
Properties \ref{Props:ChSets(1)} and Assertion \ref{As:FviUniBijection}
are kept to be true, and in the case, when the kinematic set $\fC$
is precisely visible, Properties \ref{Props:FviUni} also remain true
(but everywhere in these properties we should replace the symbol $\cZ$
by the symbol $\fC$ and the terms ``changeable set'' and ``area of
perception'' by the terms ``kinematic set'' and ``reference frame'',
correspondingly). 

\EndRmk

\section{Coordinate Transforms in Kinematic Sets.}

Let, $\fC$ be any kinematic set. For any reference frame $\lf\in\LkC$
we introduce the following denotations: 
\begin{align*}
\Mk(\lf;\fC) & :=\Tm(\lf)\times\Zk(\lf).\\
\pk{\lf}(\omega;\fC) & :=\left(\tm{\omega},\kr_{\lf}(\bs{\omega})\right)\in\Mk(\lf;\fC),\quad\omega\in\BS(\lf).
\end{align*}
The set $\Mk(\lf;\fC)$ we name as the \textbf{\emph{Minkowski set}}
of  reference frame $\lf$ in kinematic set $\fC$. The value $\pk{\lf}(\omega;\fC)$
will be named as the \textbf{\emph{Minkowski coordinates}} of the\emph{
}elementary-time state $\omega\in\BS\left(\lf\right)$\emph{ }\textbf{\emph{in
the reference frame}} $\lf$. 

In the cases, when the kinematic set $\fC$ is known in advance, we
use the denotations $\Mk(\lf)$, $\pk{\lf}(\omega)$ instead of the
denotations $\Mk(\lf;\fC)$, $\pk{\lf}(\omega;\fC)$ (correspondingly).

\BeginDef \label{Def:PeretvKoord}

Let $\fC$ be any precisely visible kinematic set and $\lf,\mf\in\LkC$
be arbitrary reference frames of~ $\fC$. 
\begin{enumerate}
\item \label{enu:RealPeretvKoord}The mapping $\PKK{\lf}{\mf}\left(\,\cdot\,;\fC\right):\BS(\lf)\mapsto\Mk(\mf)$,
represented by the formula: 
\begin{gather*}
\PKK{\lf}{\mf}(\omega;\fC)=\pk{\mf}(\unn{\lf}{\mf}{\omega}),\quad\omega\in\BS(\lf)
\end{gather*}
we name \textbf{actual coordinate transform} from $\lf$ to $\mf$.
 \\
Hence, for any $\omega\in\BS\left(\lf\right)$ the value $\PKK{\lf}{\mf}(\omega;\fC)$
may be interpreted as Minkowski coordinates of the elementary-time
state $\omega$ in the (another) reference frame $\mf\in\Lk{\fC}$. 
\item \label{enu:GlPeretvKoord(l,m)}The mapping $\widetilde{Q}:\Mk(\lf)\mapsto\Mk(\mf)$
we will name the \textbf{universal coordinate transform} from $\lf$
to $\mf$ if and only if: 

\begin{itemize}
\item $\widetilde{Q}$ is bijection (one-to-one mapping) from $\Mk(\lf)$
onto $\Mk(\mf)$.  
\item For any elementary-time state $\omega\in\BS(\lf)$ the following equality
is performed:  
\[
\PKK{\lf}{\mf}(\omega;\fC)=\widetilde{Q}\left(\pk{\lf}(\omega)\right).
\]

\end{itemize}
\item We say, that reference frames $\lf,\mf\in\Lk{\fC}$ \textbf{allow
universal coordinate transform}, if and only if at least one universal
coordinate transform $\widetilde{Q}:\Mk(\lf)\mapsto\Mk(\mf)$ from
$\lf$ to $\mf$ exists.  \\
In the case, where reference frames $\lf,\mf\in\Lk{\fC}$ allow universal
coordinate transform, we use the denotation:  
\[
\lf\gteq\limits _{\fC}\mf,
\]
 In the case, when the kinematic set $\fC$ is known in advance, we
use the abbreviated denotation~ $\lf\gteq\mf$. 
\item \label{enu:GlPeretvKoord}Indexed family of mappings $\left(\widetilde{Q}_{\mf,\lf}\right)_{\lf,\mf\in\LkC}$
will be named as the \textbf{universal coordinate transform for the
kinematic set} $\fC$  if and only if: 

\begin{itemize}
\item For arbitrary $\lf,\mf\in\LkC$ the mapping $\widetilde{Q}_{\mf,\lf}$
is universal coordinate transform from $\lf$ to $\mf$. 
\item For any $\lf,\mf,\pf\in\LkC$ and $\w\in\Mk(\lf)$ the following equalities
are true: 
\begin{gather}
\widetilde{Q}_{\lf,\lf}(\w)=\w;\qquad\widetilde{Q}_{\pf,\mf}\left(\widetilde{Q}_{\mf,\lf}(\w)\right)=\widetilde{Q}_{\pf,\lf}(\w).\label{eq:GPKconditions}
\end{gather}

\end{itemize}
\item We say, that the kinematic set $\fC$ \textbf{allows universal coordinate
transform}, if and only if there exists at least one universal coordinate
transform $\left(\widetilde{Q}_{\mf,\lf}\right)_{\lf,\mf\in\LkC}$~
for~ $\fC$. \label{enu:GlPeretvKoord1}
\end{enumerate}
\EndDef 

\BeginRmk 

In the cases, when the kinematic set $\fC$ is known in advance, we
use the abbreviated denotation $\PKK{\lf}{\mf}(\omega)$ instead of
the denotation $\PKK{\lf}{\mf}(\omega;\fC)$. 

\EndRmk 

\BeginAs[\cite{MyTmm08}] \label{As:GPK_RefFrame_properties} 

Let $\fC$ be any precisely visible kinematic set. Then: 
\begin{enumerate}
\item For an arbitrary $\lf\in\LkC$ the identity mapping on $\Mk(\lf)$:
\[
\I_{[\lf]}(\w):=\w,\qquad\w\in\Mk(\lf)
\]
 is universal coordinate transform from $\lf$ to $\lf$. 
\item If $\widetilde{Q}$ is universal coordinate transform from $\lf$
to $\mf$ ($\lf,\mf\in\LkC$), then $\arc{\widetilde{Q}}$ is universal
coordinate transform from $\mf$ to $\lf$~ (the mapping $\arc{\widetilde{Q}}$,
inverse to $\widetilde{Q}$, exists, because, according to the Definition
\ref{Def:PeretvKoord} (item 2), $\widetilde{Q}$ is bijection from
$\Mk(\lf)$ on $\Mk(\mf)$). 
\item If~ $\widetilde{Q}^{(\mf,\lf)}$ is universal coordinate transform
from $\lf$ to $\mf$, and $\widetilde{Q}^{(\pf,\mf)}$ is universal
coordinate transform from $\mf$ to $\pf$ ($\lf,\mf,\pf\in\LkC$),
then the composition of the mappings $\widetilde{Q}^{(\pf,\mf)}$
and $\widetilde{Q}^{(\mf,\lf)}$, that is the mapping: 
\begin{gather*}
\widetilde{Q}^{(\pf,\lf)}(\w)=\widetilde{Q}^{(\pf,\mf)}\left(\widetilde{Q}^{(\mf,\lf)}(\w)\right),\quad\w\in\Mk(\lf).
\end{gather*}
is universal coordinate transform from $\lf$ to $\pf$. 
\item The binary relation ~$\gteq$~ is equivalence relation on the set
~$\LkC$~ of all reference frames of ~$\fC$. 
\end{enumerate}
\EndAs

\BeginAs[\cite{MyTmm08}] \label{As:GPKkinSpaceEquivalence} 

For an arbitrary precisely visible kinematic set $\fC$ the following
propositions are equivalent: 
\begin{enumerate}
\item $\fC$ allows universal coordinate transform. 
\item For arbitrary reference frames $\lf,\mf\in\LkC$ it is true the correlation
$\lf\gteq\mf$ (that is arbitrary two reference frames $\lf,\mf\in\LkC$
allow universal coordinate transform). 
\item There exists a reference frame ~$\lf\in\LkC$~ such, that for any
reference frame ~$\mf\in\LkC$~ it is true the correlation ~$\lf\gteq\mf$. 
\end{enumerate}
\EndAs   

Let $\fC$ be any kinematic set. For arbitrary reference frame $\lf\in\LkC$
we denote:  
\begin{align}
\TR(\lf;\,\fC) & :=\left\{ \pk{\lf}(\omega)\,|\:\omega\in\BS(\lf)\right\} ;\label{eq:TrajectDef}\\
\TRd(\lf;\,\fC) & :=\Mk(\lf)\setminus\TR(\lf;\,\fC)\nonumber 
\end{align}
 (In the cases, when the kinematic set $\fC$ is known in advance,
we use the abbreviated denotations $\TR(\lf)$, $\TRd(\lf)$ instead
of the denotations $\TR(\lf;\,\fC)$, $\TRd(\lf;\,\fC)$ (correspondingly).)
 The set $\TR(\lf)$ will be named the \textbf{\emph{(general) trajectory}}
for the reference frame $\lf$, and the set  $\TRd(\lf)$ will be
named as \textbf{\emph{complement of (general) trajectory}} of the
reference frame $\lf$ in the kinematic set $\fC$.

\BeginThm[\cite{MyTmm08}] \label{Thm:GPKexistenceCriteria} 

Let $\fC$ be a precisely visible kinematic set and $\lf,\mf\in\LkC$
be any fixed reference frames of $\fC$. 

The reference frames $\lf,\mf$ allow universal coordinate transform
(i.e. $\lf\gteq\mf$) if and only if the following conditions are
satisfied:
\begin{enumerate}
\item $\card\left(\TRd(\lf)\right)=\card\left(\TRd(\mf)\right)$, ~~ where
$\card(\MM)$ means the \textbf{cardinality} of a set $\MM$. 
\item For arbitrary elementary-time states $\omega_{1},\omega_{2}\in\BS(\lf)$
the equality  $\PKK{\lf}{\mf}\left(\omega_{1}\right)=\PKK{\lf}{\mf}\left(\omega_{2}\right)$
is performed if and only if~ $\pk{\lf}\left(\omega_{1}\right)=\pk{\lf}\left(\omega_{2}\right)$. 
\end{enumerate}
\EndThm

\section{Theorems on Multi-image. }

In this section we prove the theorems on multi-image, necessary to
build mathematically strict model of kinematics of special relativity
and its extension to the kinematics that allows super-light motion
for inertial reference frames.

\subsection{Theorem on Image for Base Changeable Sets. }

\BeginDef \label{Def:EPrj}

The ordered triple $\left(\TT,\cX,U\right)$ will be referred to as
\textbf{evolution projector} for base changeable set $\BBB$ if and
only if: 

1. $\TT=\left(\T,\leq\right)$~ is linearly ordered set; 

2. $\text{\ensuremath{\cX}\ }$ is any set; 

3. $U$~ is a mapping from $\BSB$ into $\T\times\cX$~ ($U:\BSB\mapsto\T\times\cX$). 

\EndDef 

\BeginDef[\cite{MyTmm04}]\label{Def:UnitedByFate}

Let $\BBB$ be any base changeable set. We will say, that elementary-time
states $\omega_{1},\omega_{2}\in\BSB$\emph{ }are \textbf{\emph{united
by fate}} in $\BBB$ if and only if at least one of the conditions
$\omega_{2}\fff\omega_{1}$ or $\omega_{1}\fff\omega_{2}$ is satisfied. 

\EndDef 

\BeginThm[\cite{MyTmm06,MyTmm03}] \label{Thm:Img}

Let $\left(\TT,\cX,U\right)$ be any evolution projector for base
changeable set $\BBB$. Then there exist only one base changeable
set $U\left[\BBB,\TT\right]$, satisfying the following conditions: 
\begin{enumerate}
\item $\TM\left(U\left[\BBB,\TT\right]\right)=\TT$;{\large \vspace{-1mm}
}{\large \par}
\item $\BS(U\left[\BBB,\TT\right])=U(\BSB)=\,$$\left\{ U(\omega)\,|\:\omega\in\BSB\right\} $;{\large \vspace{-1mm}
}{\large \par}
\item Let $\widetilde{\omega}_{1},\widetilde{\omega}_{2}\in\BS(U\left[\BBB,\TT\right])$
and $\tm{\widetilde{\omega}_{1}}\neq\tm{\widetilde{\omega}_{2}}$.
Then $\widetilde{\omega}_{1}$ and $\widetilde{\omega}_{2}$ are united
by fate in $U\left[\BBB,\TT\right]$ if and only if, there exist united
by fate in $\BBB$ elementary-time states $\omega_{1},\omega_{2}\in\BSB$
such, that $\widetilde{\omega}_{1}=U\left(\omega_{1}\right)$, $\widetilde{\omega}_{2}=U\left(\omega_{2}\right)$. 
\end{enumerate}
\EndThm

\BeginRmk 

In the case, when $\TT=\TMB$ we use the denotation $U\left[\BBB\right]$
instead of the denotation $U\left[\BBB,\TT\right]$: 
\[
U\left[\BBB\right]:=U\left[\BBB,\TMB\right].
\]

\EndRmk 

\BeginRmk \label{Rmk:IdBmmImg} 

Let $\BBB$ be any base changeable set and $\I_{\BSB}:\BSB\mapsto\TmB\times\BsB$
be the mapping, given by the formula: $\I_{\BSB}(\omega)=\omega$~
$\left(\omega\in\BSB\right)$. Then the triple $\left(\TMB,\BsB,\I_{\BSB}\right)$,
is, apparently, evolution projector for $\BBB$. Moreover, if we substitute
$\TMB$ and $\BBB$ into Theorem \ref{Thm:Img} instead of $\TT$
and $U\left[\BBB,\TT\right]$ (correspondingly), we can see, that
all conditions of this Theorem are satisfied. Hence for the identity
mapping $\I_{\BSB}$ (on $\BSB$), we obtain: 
\[
\I_{\BSB}\left[\BBB\right]=\BBB.
\]

\EndRmk

\subsection{Theorem on Multi-image for Changeable Sets. }

Further $\Rg(U)$ will mean the \textbf{\emph{range}} of (arbitrary)
mapping $U$. 

\BeginDef  \label{Def:EmultiPrj} 
\begin{enumerate}
\item The evolution projector $\left(\TT,\cX,U\right)$ (where $\TT=\left(\T,\leq\right)$)
for base changeable set $\BBB$ will be named \textbf{bijective} if
and only if the mapping $U$ is bijection from $\BSB$ onto the set
$\Rg(U)\subseteq\T\times\cX$.\label{enu:BijectEprj} 
\item Any indexed family $\fP=\left(\left(\TT_{\alpha},\cX_{\alpha},U_{\alpha}\right)\,|\:\alpha\in\AAA\right)$
(where $\AAA\neq\emptyset$) of bijective evolution projectors for
base changeable set we name \textbf{evolution multi-projector} for
$\BBB$.\label{enu:EmultiPrj} 
\end{enumerate}
\EndDef  

\BeginThm  \label{Thm:MultiImg}

Let, $\fP=\left(\left(\TT_{\alpha},\cX_{\alpha},U_{\alpha}\right)\,|\:\alpha\in\AAA\right)$
be evolution multi-projector for base changeable set $\BBB$. Then:
\begin{description}
\item [{A.}] Only one changeable set $\cZ$ exists, satisfying the following
conditions: \end{description}
\begin{enumerate}
\item $\LkZ=\left\{ \left(\alpha,U_{\alpha}\left[\BBB,\TT_{\alpha}\right]\right)\,|\;\alpha\in\AAA\right\} $.\label{enu:Zim:Lk}
\item For any areas of perception $\lf=\left(\alpha,U_{\alpha}\left[\BBB,\TT_{\alpha}\right]\right)\in\LkZ$,
$\mf=\left(\beta,U_{\beta}\left[\BBB,\TT_{\beta}\right]\right)\in\LkZ$
($\alpha,\beta\in\AAA$) and any set $A\subseteq\BS(\lf)=U_{\alpha}(\BSB)$
the following equality holds: 
\begin{gather*}
\un{\lf,\cZ}{\mf}A=U_{\beta}\left(\arc{U_{\alpha}}(A)\right)=\left\{ U_{\beta}\left(\arc{U_{\alpha}}(\omega)\right)\,|\:\omega\in A\right\} ,
\end{gather*}
 where $\arc{U_{\alpha}}$ is the mapping, \textbf{inverse} to $U_{\alpha}$.
\label{enu:Zim:Unif} \end{enumerate}
\begin{description}
\item [{B.}] Changeable set $\cZ$, satisfying the conditions \ref{enu:Zim:Lk},~\ref{enu:Zim:Unif}
is precisely visible. 
\end{description}
\EndThm 

\BeginRmk 

Suppose, that a changeable set $\cZ$ satisfies condition \ref{enu:Zim:Lk}
of Theorem \ref{Thm:MultiImg}. Then for any area of perception $\lf=\left(\alpha,U_{\alpha}\left[\BBB,\TT_{\alpha}\right]\right)\in\LkZ$,
according to Property \ref{Props:ChSets(1)}(\ref{enu:PropChSets:Oll}),
we have, $\ind{\lf}=\alpha$, $\oll=U_{\alpha}\left[\BBB,\TT_{\alpha}\right]$,
and hence, $\BS(\lf)=\BS\left(\oll\right)=\BS\left(U_{\alpha}\left[\BBB,\TT_{\alpha}\right]\right)$.
Therefore, by Theorem \ref{Thm:Img}, $\BS(\lf)=U_{\alpha}(\BSB)$.
Thus, the condition \ref{enu:Zim:Unif} of Theorem \ref{Thm:MultiImg}
is correctly formulated. 

\EndRmk  

\BeginProof[Proof of Theorem \ref{Thm:MultiImg}] 

Let, $\fP=\left(\left(\TT_{\alpha},\cX_{\alpha},U_{\alpha}\right)\,|\:\alpha\in\AAA\right)$
be evolution multi-projector for base changeable set $\BBB$. 

\textbf{A.} By Definition \ref{Def:EmultiPrj}, for any $\alpha\in\AAA$
the triple $\left(\TT_{\alpha},\cX_{\alpha},U_{\alpha}\right)$, is
a bijective evolution projector for $\BBB$. In accordance with Theorem
\ref{Thm:Img}, we put:
\[
\BBB_{\alpha}:=U_{\alpha}\left[\BBB,\TT_{\alpha}\right]\quad\left(\alpha\in\AAA\right).
\]
 Since $\left(\TT_{\alpha},\cX_{\alpha},U_{\alpha}\right)$ is a bijective
evolution projector, then, by the Definition \ref{Def:EmultiPrj},
the mapping $U_{\alpha}$ is one-to-one correspondence. Hence, the
inverse mapping $\arc{U_{\alpha}}$ exists (for all $\alpha\in\AAA$). 

For any indexes $\alpha,\beta\in\AAA$ and any set $A\subseteq\BSBB{\alpha}$
we denote: 
\begin{equation}
\fU_{\beta,\alpha}A:=U_{\beta}\left(\arc{U_{\alpha}}(A)\right)\label{eq:MultiIm:UnificDef}
\end{equation}
(note, that, by Theorem \ref{Thm:Img}, $\BSBB{\alpha}=U_{\alpha}\left(\BSB\right)$).
Hence, $\fU_{\beta,\alpha}$ is the mapping from $2^{\BSBB{\alpha}}$
into $2^{\BSBB{\beta}}=2^{U_{\beta}\left(\BSB\right)}$.

It is easy to verify, that the family of mappings $\left(\fU_{\beta,\alpha}\,|\:\alpha,\beta\in\AAA\right)$
possesses the following properties: 
\begin{enumerate}
\item $\fU_{\alpha,\alpha}\, A=A$ ($\alpha\in\AAA$, $A\subseteq\BSBB{\alpha}$);{\large \vspace{-1mm}
}{\large \par}
\item If $\alpha,\beta\in\AAA$ and $A\subseteq B\subseteq\BSBB{\alpha}$,
then $\fU_{\beta,\alpha}A\subseteq\fU_{\beta,\alpha}B$. {\large \vspace{-1mm}
}{\large \par}
\item If $\alpha,\beta,\gamma\in\AAA$ and $A\subseteq\BSBB{\alpha}$, then{\large{}
\vspace{-1mm}
} 
\begin{equation}
\fU_{\gamma,\beta}\fU_{\beta,\alpha}A=\fU_{\gamma,\alpha}A.\label{eq:MultiIm:Fvi}
\end{equation}

\end{enumerate}
Therefore, by \cite[Definition 9.6]{MyTmm03} or \cite[Definition 3.1]{MyTmm06},
the triple: 
\begin{gather*}
\cZ=\left(\AAA,\vcB,\vfU\right),\;\text{where}\quad\vcB=\left(\BBB_{\alpha}\,|\:\alpha\in\AAA\right);\:\vfU=\left(\fU_{\beta,\alpha}\,|\:\alpha,\beta\in\AAA\right).
\end{gather*}
 is changeable set. Herewith, according to system of denotations for
the changeable sets theory, accepted in \cite{MyTmm03},~\cite{MyTmm06}:
\begin{gather}
\LkZ=\left\{ \left(\alpha,\BBB_{\alpha}\right)\,|\:\alpha\in\AAA\right\} =\left\{ \left(\alpha,U_{\alpha}\left[\BBB,\TT_{\alpha}\right]\right)\,|\;\alpha\in\AAA\right\} ,\label{eq:MultiIm:Lk}
\end{gather}
 and for arbitrary areas of perception $\lf=\left(\alpha,U_{\alpha}\left[\BBB,\TT_{\alpha}\right]\right)\in\LkZ$,
$\mf=\left(\beta,U_{\beta}\left[\BBB,\TT_{\beta}\right]\right)\in\LkZ$
(where $\alpha,\beta\in\AAA$)  and for any set $A\subseteq\BS(\lf)=\BS\left(U_{\alpha}\left[\BBB,\TT_{\alpha}\right]\right)=U_{\alpha}(\BSB)$,
by (\ref{eq:MultiIm:UnificDef}) we obtain: 
\begin{gather}
\un{\lf,\,\cZ}{\mf}A=\fU_{\beta,\alpha}A=U_{\beta}\left(\arc{U_{\alpha}}(A)\right).\label{eq:MultiIm:Unific}
\end{gather}
From (\ref{eq:MultiIm:Lk}) and (\ref{eq:MultiIm:Unific}) it follows,
that the changeable set $\cZ$ satisfies conditions \ref{enu:Zim:Lk},\ref{enu:Zim:Unif}
of Theorem \ref{Thm:MultiImg}. 

Suppose, that the changeable set $\cZ_{1}$ also satisfies conditions
\ref{enu:Zim:Lk},\ref{enu:Zim:Unif} of Theorem \ref{Thm:MultiImg}.
Then, by the condition \ref{enu:Zim:Lk}, $\LkZ=\Lk{\cZ_{1}}$. Also,
by the condition \ref{enu:Zim:Unif}, for arbitrary areas of perception
$\lf,\mf\in\LkZ=\Lk{\cZ_{1}}$ it is true the equality: $\un{\lf,\cZ}{\mf}=\un{\lf,\cZ_{1}}{\mf}$.
Hence, by Assertion \ref{As:Z1=00003DZ2_(Lk,unif)}, we get $\cZ=\cZ_{1}$.~
Thus, changeable set, satisfying the conditions \ref{enu:Zim:Lk},\ref{enu:Zim:Unif}
of Theorem \ref{Thm:MultiImg} is unique. 

\textbf{B. }Using the equalities (\ref{eq:MultiIm:Unific}) and (\ref{eq:MultiIm:Fvi}),
for arbitrary areas of perception  $\lf,\mf,\pf\in\LkZ$ of kind $\lf=\left(\alpha,U_{\alpha}\left[\BBB,\TT_{\alpha}\right]\right)$,
$\mf=\left(\beta,U_{\beta}\left[\BBB,\TT_{\beta}\right]\right)$,
$\pf=\left(\gamma,U_{\gamma}\left[\BBB,\TT_{\gamma}\right]\right)$
($\alpha,\beta,\gamma\in\AAA$) we get: 
\begin{gather*}
\un{\mf}{\pf}\un{\lf}{\mf}A=\fU_{\gamma,\beta}\fU_{\beta,\alpha}A=\fU_{\gamma,\alpha}A=\un{\lf}{\pf}A\quad\left(A\subseteq\BS(\lf)\right).
\end{gather*}
Thus, by \cite[Theorem 5.1]{MyTmm04} or \cite[Theorem 11.1]{MyTmm03},
the changeable set $\cZ$ is precisely visible.   \EndProof  

\BeginDef \label{Def:MultiImg} Let $\fP=\left(\left(\TT_{\alpha},\cX_{\alpha},U_{\alpha}\right)\,|\:\alpha\in\AAA\right)$
be evolution multi-projector for base changeable set $\BBB$. Changeable
set $\cZ$, satisfying conditions \ref{enu:Zim:Lk},\ref{enu:Zim:Unif}
of Theorem \ref{Thm:MultiImg} will be referred to as \textbf{evolution
multi-image} of base changeable set $\BBB$ relatively the evolution
multi-projector $\fP$. This evolution multi-image will be denoted
by $\Zimm{\fP,\BBB}$: 
\[
\Zimm{\fP,\BBB}:=\cZ.
\]

\EndDef 

From the theorems \ref{Thm:MultiImg} and \ref{Thm:Img}, taking into
account Property \ref{Props:BMM}(\ref{enu:BMMProp(Bs)}), Property
\ref{Props:ChSets(0)}(1) and Remark \ref{Rmk:BChSet_to_lf}, we immediately
deduce the following properties of multi-image for base changeable
set. 

\BeginProp  \label{Props:Zim:1}

Let $\fP=\left(\left(\TT_{\alpha},\cX_{\alpha},U_{\alpha}\right)\,|\:\alpha\in\AAA\right)$,
where $\TT_{\alpha}=\left(\T_{\alpha},\leq_{\alpha}\right)$ ($\alpha\in\AAA$)
be an evolution multi-projector for base changeable set $\BBB$ and
$\cZ=\Zimm{\fP,\BBB}$. Then: 
\begin{enumerate}
\item $\LkZ=\left\{ \left(\alpha,U_{\alpha}\left[\BBB,\TT_{\alpha}\right]\right)\,|\:\alpha\in\AAA\right\} $.
\item $\IndZ=\AAA$. 
\item For any area of perception $\lf=\left(\alpha,U_{\alpha}\left[\BBB,\TT_{\alpha}\right]\right)$
the following equalities hold: \label{enu:Props:Zim:1:BasEqs} 
\begin{align*}
\BS(\lf) & =U_{\alpha}\left(\BSB\right)=\left\{ U_{\alpha}(\omega)\,|\:\omega\in\BSB\right\} ;\\
\Bs(\lf) & =\left\{ \bs{U_{\alpha}(\omega)}\,|\:\omega\in\BSB\right\} ;\\
\TM(\lf) & =\TT_{\alpha};\quad\Tm(\lf)=\T_{\alpha}.
\end{align*}

\item Let, $\lf=\left(\alpha,U_{\alpha}\left[\BBB,\TT_{\alpha}\right]\right)\in\LkZ$,
where $\alpha\in\AAA$.  Suppose, that $\widetilde{\omega}_{1},\widetilde{\omega}_{2}\in\BS(\lf)$
and $\tm{\widetilde{\omega}_{1}}\neq\tm{\widetilde{\omega}_{2}}$.
Then $\widetilde{\omega}_{1}$ and $\widetilde{\omega}_{2}$ are united
by fate in $\lf$ if and only if there exist united by fate in $\BBB$
elementary-time states $\omega_{1},\omega_{2}\in\BSB$ such, that
$\widetilde{\omega}_{1}=U_{\alpha}\left(\omega_{1}\right)$, $\widetilde{\omega}_{2}=U_{\alpha}\left(\omega_{2}\right)$. 
\end{enumerate}
\EndProp 

Let $\BBB$ be any base changeable set, and $X$ be any set, containing
$\BsB$ ($\BsB\subseteq X$). Recall (\cite[Example 3.2]{MyTmm06},
\cite[Example 10.2]{MyTmm03}), that any set $\bbU$ of bijections,
defined on $\TmB\times X$: 
\[
U:\TmB\times X\longleftrightarrow\TmB\times X\quad(U\in\bbU)
\]
is named \textbf{\emph{transforming set of bijections}}\emph{ }relatively
the base changeable set $\BBB$ on $X$. 

Let $\bbU$ be transforming set of bijections relatively $\BBB$ on
$X$. Then, by the Definition \ref{Def:EPrj}, any mapping $U\in\bbU$
generates the evolution projector, $\left(\TMB,X,U_{\upharpoonright\BSB}\right),$
where $U_{\upharpoonright\BSB}$ is the restriction of the mapping
$U$ onto the set $\BSB\subseteq\TmB\times X$ (henceforth, where
it does not cause confusion, we identify the mapping $U_{\upharpoonright\BSB}$
with the mapping $U$; under this identifying, we can consider, that
$\left(\TMB,X,U_{\upharpoonright\BSB}\right)=\left(\TMB,X,U\right)$).
Hence, the indexed family: 
\[
\fP_{\BBB}\left[\bbU\right]=\left(\left(\TMB,X,U\right)\,|\: U\in\bbU\right)
\]
 is evolution multi-projector for $\BBB$. In this particular case
we have the changeable set: 
\[
\Zim{\bbU,\BBB}=\Zimm{\fP_{\BBB}\left[\bbU\right],\BBB},
\]
that is multi-figurative image of base changeable set $\BBB$ relatively
the transforming set of bijections $\bbU$ (in the sense of \cite{MyTmm06,MyTmm03}). 

From the item B of Theorem \ref{Thm:MultiImg} it follows, that the
changeable sets of kind $\Zimm{\fP,\BBB}$ and $\Zim{\bbU,\BBB}$
are precisely visible. Therefore, we deliver the following corollary
of Theorem \ref{Thm:MultiImg}: 

\BeginNasl \label{Nasl:Zim<uni!>} 

If $\cZ=\Zimm{\fP,\BBB}$, where $\fP=\left(\left(\TT_{\alpha},\cX_{\alpha},U_{\alpha}\right)\,|\:\alpha\in\AAA\right)$,
then for any areas of perception $\lf=\left(\alpha,U_{\alpha}\left[\BBB,\TT_{\alpha}\right]\right)\in\LkZ$,
$\mf=\left(\beta,U_{\beta}\left[\BBB,\TT_{\beta}\right]\right)\in\LkZ$
($\alpha,\beta\in\AAA$) the following equality is performed: 
\[
\unn{\lf,\,\cZ}{\mf}{\omega}=U_{\beta}\left(\arc{U_{\alpha}}(\omega)\right)\quad\left(\omega\in\BS(\lf)=U_{\alpha}\left(\BSB\right)\right).
\]

\EndNasl

\subsection{Theorem on Multi-image for Kinematic Sets. }

\BeginDef \label{Def:KnPrj}\label{Def:KnMultiPrj}~
\begin{enumerate}
\item The ordered composition of five sets $\left(\TT,\cX,U,\kpQ,k\right)$
will be named as bijective \textbf{kinematic projector} for base changeable
set $\BBB$ if and only if: 

\begin{description}
\item [{1.1.}] $\left(\TT,\cX,U\right)$~ is bijective evolution projector
for $\BBB$. 
\item [{1.2.}] $\kpQ$~ is a coordinate space. 
\item [{1.3.}] $k$ is a mapping from $\cX$ into $\ZkQ$. 
\end{description}
\item Any indexed family~ $\fP=\left(\left(\TT_{\alpha},\cX_{\alpha},U_{\alpha},\kpQ_{\alpha},k_{\alpha}\right)\,|\:\alpha\in\AAA\right)$~
(where $\AAA\neq\emptyset$) of bijective kinematic projectors for
base changeable set~ $\BBB$ ~we name \textbf{kinematic multi-projector}
for $\BBB$. 
\end{enumerate}
\EndDef 

\BeginRmk 

Henceforward we will consider only bijective kinematic projectors.
That is why we will use the term ``kinematic projector'' instead of
the term ``bijective kinematic projector''.

\EndRmk 

Let $\fP=\left(\left(\TT_{\alpha},\cX_{\alpha},U_{\alpha},\kpQ_{\alpha},k_{\alpha}\right)\,|\:\alpha\in\AAA\right)$
be any kinematic multi-projector for $\BBB$. Denote:
\[
\fP^{\left[e\right]}:=\left(\left(\TT_{\alpha},\cX_{\alpha},U_{\alpha}\right)\,|\:\alpha\in\AAA\right).
\]
By the definitions \ref{Def:KnMultiPrj} and \ref{Def:EmultiPrj},
$\fP^{\left[e\right]}$ is evolution multi-projector for $\BBB$. 

\BeginThm \label{Thm:MultiImg(KM)} 

Let $\fP=\left(\left(\TT_{\alpha},\cX_{\alpha},U_{\alpha},\kpQ_{\alpha},k_{\alpha}\right)\,|\:\alpha\in\AAA\right)$
be a kinematic multi-projector for a base changeable set $\BBB$.
Then:

\noindent \textbf{A)} Only one kinematic set $\fC$ exists, satisfying
the following conditions: 
\begin{enumerate}
\item $\BBE\left(\fC\right)=\Zimm{\fP^{\left[e\right]},\BBB}$. 
\item For any reference frame $\lf=\left(\alpha,U_{\alpha}\left[\BBB,\TT_{\alpha}\right]\right)\in\LkC$
(where $\alpha\in\AAA$) the following equalities are performed:

\begin{center}
\em{\textbf{2.1)} $\BG(\lf)=\kpQ_{\alpha}$; ~ ~ ~ \textbf{2.2)}
$\kr_{\lf}(x)=k_{\alpha}(x)\quad(x\in\Bs(\lf))$.}
\par\end{center}

\end{enumerate}
\textbf{B) }Kinematic set $\fC$, satisfying the conditions 1,2 is
precisely visible. 

\EndThm 

\BeginProof  

Let $\fP=\left(\left(\TT_{\alpha},\cX_{\alpha},U_{\alpha},\kpQ_{\alpha},k_{\alpha}\right)\,|\:\alpha\in\AAA\right)$~
(where $\TT_{\alpha}=\left(\T_{\alpha},\leq_{\alpha}\right)$, $\alpha\in\AAA$)
be a kinematic multi-projector for $\BBB$.

\textbf{A)} Put: 
\[
\cZ:=\Zimm{\fP^{\left[e\right]},\BBB}.
\]
 Then, according to Theorem \ref{Thm:MultiImg}: 
\[
\LkZ=\left\{ \left(\alpha,U_{\alpha}\left[\BBB,\TT_{\alpha}\right]\right)\,|\right.\left.\;\alpha\in\AAA\right\} .
\]
 Consider any fixed area of perception $\lf=\left(\alpha,U_{\alpha}\left[\BBB,\TT_{\alpha}\right]\right)\in\LkZ$
(where $\alpha\in\AAA$). Denote: 
\[
\kpQ^{(\lf)}:=\kpQ_{\alpha}.
\]

The ordered five-elements composition $\left(\TT_{\alpha},\cX_{\alpha},U_{\alpha},\kpQ_{\alpha},k_{\alpha}\right)$
is a kinematic projector. Hence, by the Definition \ref{Def:KnPrj},
the triple $\left(\TT_{\alpha},\cX_{\alpha},U_{\alpha}\right)=\left(\left(\T_{\alpha},\leq_{\alpha}\right),\cX_{\alpha},U_{\alpha}\right)$
is evolution projector for $\BBB$. Consequently, by the definition
of evolution projector (Definition \ref{Def:EPrj}), $U_{\alpha}$
is the mapping of kind $U_{\alpha}:\BSB\mapsto\T_{\alpha}\times\cX_{\alpha}$.
Therefore, by Property \ref{Props:Zim:1}(\ref{enu:Props:Zim:1:BasEqs}),
we obtain:  
\[
\Bs(\lf)=\left\{ \bs{U_{\alpha}(\omega)}\,|\:\omega\in\BSB\right\} \subseteq\cX_{\alpha}.
\]

For an arbitrary $x\in\Bs(\lf)$ we denote: 
\[
k^{(\lf)}(x):=k_{\alpha}(x).
\]
 According to the definition of a kinematic projector (Definition
\ref{Def:KnPrj}) $k_{\alpha}$ is the mapping from $\cX_{\alpha}$
into $\Zk\left(\kpQ_{\alpha}\right)=\Zk\left(\kpQ^{(\lf)}\right)$.
Hence, $k^{(\lf)}$ is the mapping from $\Bs(\lf)$ into $\Zk\left(\kpQ^{(\lf)}\right)$. 

Hence, by the Definition \ref{Def:KinSets} (item \ref{enu:Def:KSets:General}),
the pair 
\begin{equation}
\fC=\left(\cZ,\,\left(\left(\kpQ^{(\lf)},\, k^{(\lf)}\right)\,|\:\lf\in\LkZ\right)\right)\label{eq:KinSetCDef}
\end{equation}
is a kinematic set. Herewith, taking into account the system of denotations,
accepted in the subsection \ref{sub:fKMMDenotations}, we get: 
\[
\BBE(\fC)=\cZ=\Zimm{\fP^{\left[e\right]}},
\]
and for any reference frame $\lf=\left(\alpha,U_{\alpha}\left[\BBB,\TT_{\alpha}\right]\right)\in\LkC$,
where $\alpha\in\AAA$ we have: 
\begin{align*}
\BG(\lf) & =\kpQ^{(\lf)}=\kpQ_{\alpha};\\
\kr_{\lf}(x) & =k^{(\lf)}(x)=k_{\alpha}(x)\quad(x\in\Bs(\lf)).
\end{align*}

Thus, the kinematic set $\fC$ satisfies the conditions 1,2 from the
item \textbf{A)} of Theorem \ref{Thm:MultiImg(KM)}. 

Now, we are going to prove, that kinematic set $\fC$, satisfying
the conditions 1,2 from the item \textbf{A)} of Theorem \ref{Thm:MultiImg(KM)}
is unique. Assume, that the kinematic set $\fC_{1}$ also satisfies
the conditions 1,2 from the item \textbf{A)} of Theorem \ref{Thm:MultiImg(KM)}.
Then, by the condition 1, $\BBE\left(\fC\right)=\cZ=\BBE\left(\fC_{1}\right)$.
Hence, 
\[
\LkC=\LkZ=\Lk{\fC_{1}},
\]
moreover, for any reference frames $\lf,\mf\in\LkC=\Lk{\fC_{1}}$
we have: 
\begin{gather*}
\un{\lf,\fC}{\mf}=\un{\lf,\cZ}{\mf}=\un{\lf,\fC_{1}}{\mf}.
\end{gather*}
Further, by the condition 2, for any reference frame $\lf=\left(\alpha,U_{\alpha}\left[\BBB,\TT_{\alpha}\right]\right)\in\LkC=\Lk{\fC_{1}}$
we deliver: 
\begin{align*}
\BG\left(\lf;\,\fC\right) & =\kpQ_{\alpha}=\BG\left(\lf;\,\fC_{1}\right);\qquad\kr_{\lf}\left(x,\,\fC\right)=k_{\alpha}(x)=\kr_{\lf}\left(x,\,\fC_{1}\right)\;\:\left(x\in\Bs(\lf)\right).
\end{align*}
Therefore, by Assertion \ref{As:Nasl:C1=00003DC2_(Lk,BG(l),ql,unif)},
$\fC=\fC_{1}$. 

\textbf{B)} Since the changeable set $\cZ$ is precisely visible,
then the kinematic set $\fC$, represented, by the formula (\ref{eq:KinSetCDef}),
according to the item d) of the subsection \ref{sub:fKMMDenotations},
also is precisely visible.      \EndProof  

\BeginDef \label{Def:MultiImg(KM)}

Let, $\fP=\left(\left(\TT_{\alpha},\cX_{\alpha},U_{\alpha},\kpQ_{\alpha},k_{\alpha}\right)\,|\:\alpha\in\AAA\right)$
be a kinematic multi-projector for a base changeable set $\BBB$.
The kinematic set $\fC$, satisfying the conditions 1,2 of Theorem
\ref{Thm:MultiImg(KM)} will be named as \textbf{kinematic multi-image}
of base changeable set $\BBB$ relatively the kinematic multi-projector
$\fP$. This kinematic set will be denoted via $\Kimm{\fP,\BBB}$:
\[
\Kimm{\fP,\BBB}:=\fC.
\]

\EndDef 

Applying Properties \ref{Props:Zim:1}, Corollary \ref{Nasl:Zim<uni!>}
and Theorem \ref{Thm:MultiImg(KM)}, we obtain the following properties
for kinematic multi-image of base changeable set. 

\BeginProp  \label{Props:Kim:1}

Let, $\fP=\left(\left(\TT_{\alpha},\cX_{\alpha},U_{\alpha},\kpQ_{\alpha},k_{\alpha}\right)\,|\:\alpha\in\AAA\right)$
be a kinematic multi-projector for $\BBB$ (where $\TT_{\alpha}=\left(\T_{\alpha},\leq_{\alpha}\right)$,
$\alpha\in\AAA$). Suppose, that $\fC=\Kimm{\fP,\BBB}$. Then: 
\begin{enumerate}
\item $\LkC=\left\{ \left(\alpha,U_{\alpha}\left[\BBB,\TT_{\alpha}\right]\right)\,|\:\alpha\in\AAA\right\} $.
\label{enu:PropKim:1:Lk}
\item $\Ind{\fC}=\AAA$. 
\item For any reference frame $\lf=\left(\alpha,U_{\alpha}\left[\BBB,\TT_{\alpha}\right]\right)$
the following equalities hold: \label{enu:PropKim:1:FrameComponents}
\begin{align*}
\BS(\lf) & =U_{\alpha}\left(\BSB\right)=\left\{ U_{\alpha}(\omega)\,|\:\omega\in\BSB\right\} ;\\
\Bs(\lf) & =\left\{ \bs{U_{\alpha}(\omega)}\,|\:\omega\in\BSB\right\} ;\\
\TM(\lf) & =\TT_{\alpha};\quad\Tm(\lf)=\T_{\alpha};\\
\Zk(\lf) & =\Zk\left(\BG(\lf)\right)=\Zk\left(\kpQ_{\alpha}\right);\\
\Mk(\lf) & =\Tm(\lf)\times\Zk(\lf)=\T_{\alpha}\times\Zk\left(\kpQ_{\alpha}\right);\\
\kr_{\lf}(x) & =k_{\alpha}(x)\quad(x\in\Bs(\lf));\\
\pk{\lf}(\omega) & =\left(\tm{\omega},\kr_{\lf}(\bs{\omega})\right)=\left(\tm{\omega},k_{\alpha}(\bs{\omega})\right)\qquad\left(\omega\in\BS(\lf)\right).
\end{align*}

\item Let, $\lf=\left(\alpha,U_{\alpha}\left[\BBB,\TT_{\alpha}\right]\right)\in\LkC$,
where $\alpha\in\AAA$.  Suppose, that $\widetilde{\omega}_{1},\widetilde{\omega}_{2}\in\BS(\lf)$
and $\tm{\widetilde{\omega}_{1}}\neq\tm{\widetilde{\omega}_{2}}$.
Then $\widetilde{\omega}_{1}$ and~ $\widetilde{\omega}_{2}$ are
united by fate in $\lf$ if and only if there exist united by fate
in $\BBB$ elementary-time states $\omega_{1},\omega_{2}\in\BSB$
such, that $\widetilde{\omega}_{1}=U_{\alpha}\left(\omega_{1}\right)$,
$\widetilde{\omega}_{2}=U_{\alpha}\left(\omega_{2}\right)$. 
\item For any reference frames~ $\lf=\left(\alpha,U_{\alpha}\left[\BBB,\TT_{\alpha}\right]\right)\in\LkC$,
$\mf=\left(\beta,U_{\beta}\left[\BBB,\TT_{\beta}\right]\right)\in\LkC$
($\alpha,\beta\in\AAA$) the following equality holds:\label{enu:PropKim:1:Unif!}
\[
\unn{\lf,\,\fC}{\mf}{\omega}=U_{\beta}\left(\arc{U_{\alpha}}(\omega)\right)\quad\left(\omega\in\BS(\lf)=U_{\alpha}\left(\BSB\right)\right).
\]

\end{enumerate}
\EndProp

\section{Kinematic Sets, Generated by Special Relativity and its Tachyon Extensions. }

Let, $\kpQ$ be a coordinate space, $\BBB$ be a base changeable set
such, that $\BsB\subseteq\ZkQ$ (such base changeable set $\BBB$
exists, because, for example, we may put $\BBB:=\At\left(\TT,\cR\right)$,
where $\cR$ is a system of abstract trajectories from the linear
ordered set $\TT$ to a set $\mathbf{M}\subseteq\ZkQ$, where the
definition of $\At\left(\TT,\cR\right)$ can be found in \cite{MyTmm05(YMJ),MyTmm03}).
Let $\bbU$ be any transforming set of bijections relatively the $\BBB$
on $\ZkQ$. Then, any mapping $\mathbf{U}\in\bbU$ is the mapping
of kind, $\mathbf{U}:\TmB\times\ZkQ\longleftrightarrow\TmB\times\ZkQ$,
where $\BSB\subseteq\TmB\times\BsB\subseteq\TmB\times\ZkQ$. Hence,
the set of bijections $\bbU$ generates the kinematic multi-projector
$\bbUx:=\left(\left(\TMB,\ZkQ,\mathbf{U},\kpQ,\I_{\ZkQ}\right)\,|\:\mathbf{U}\in\bbU\right)$
for $\BBB$, where $\I_{\ZkQ}$ is the identity mapping on $\ZkQ$.
 Denote: 
\[
\Kim{\bbU,\BBB,\kpQ}:=\Kimm{\bbUx,\BBB}.
\]

\BeginThm  \label{Thm:KimUniversal} 

The kinematic set $\fC=\Kim{\bbU,\BBB,\kpQ}$ allows universal coordinate
transform. Moreover, $\LkC=\left(\left(\mathbf{U},\mathbf{U}\left[\BBB\right]\right)\,|\:\mathbf{U}\in\bbU\right)$,
and the system of mappings $\left(\widetilde{Q}_{\mf,\lf}\right)_{\lf,\mf\in\LkC}$:
\vspace{-3mm}

\begin{align}
\widetilde{Q}_{\mf,\lf}(\w) & =\mathbf{V}\left(\arc{\mathbf{U}}(\w)\right),\quad\w\in\Mk(\lf)=\TmB\times\ZkQ\label{eq:Kim:UniversalPk}\\
 & \qquad\left(\,\lf=\left(\mathbf{U},\mathbf{U}\left[\BBB\right]\right)\in\LkC,\quad\mf=\left(\mathbf{V},\mathbf{V}\left[\BBB\right]\right)\in\LkC\,\right)\nonumber 
\end{align}
 is universal coordinate transform for $\fC$. 

\EndThm 

\BeginProof 

Let, $\kpQ$ be a coordinate space and $\bbU$ be transforming set
of bijections relatively the base changeable set $\BBB$ ($\BsB\subseteq\ZkQ$)
on $\ZkQ$. Denote $\fC=\Kim{\bbU,\BBB,\kpQ}$. Then, $\fC=\Kimm{\bbUx,\BBB}$,
where $\bbUx=\left(\left(\TMB,\ZkQ,\mathbf{U},\kpQ,\I_{\ZkQ}\right)\,|\:\mathbf{U}\in\bbU\right)$.
Hence, according to Property \ref{Props:Kim:1}(\ref{enu:PropKim:1:Lk}),
$\LkC=\left\{ \left(\mathbf{U},\mathbf{U}\left[\BBB\right]\right)\,|\:\mathbf{U}\in\bbU\right\} $.
And, by Property \ref{Props:Kim:1}(\ref{enu:PropKim:1:FrameComponents}),
for an arbitrary reference frame $\lf=\left(\mathbf{U},\mathbf{U}\left[\BBB\right]\right)\in\LkC$
we have: $\Bs(\lf)=\left\{ \bs{\mathbf{U}(\omega)}\,|\:\omega\in\BSB\right\} \subseteq\ZkQ$.
Herewith, by Theorem  \ref{Thm:MultiImg(KM)}, $\kr_{\lf}\left(x,\,\fC\right)=x$~
$(\forall\, x\in\Bs(\lf))$. Hence: 
\[
\pk{\lf}(\omega;\fC)=\left(\tm{\omega},\kr_{\lf}(\bs{\omega})\right)=\left(\tm{\omega},\bs{\omega}\right)=\omega\quad\left(\lf\in\LkC,\;\omega\in\BS(\lf)\right).
\]
 Using the last equality and Property \ref{Props:Kim:1}(\ref{enu:PropKim:1:Unif!}),
for arbitrary reference frames $\lf=\left(\mathbf{U},\mathbf{U}\left[\BBB\right]\right)\in\LkC$,
$\mf=\left(\mathbf{V},\mathbf{V}\left[\BBB\right]\right)\in\LkC$~
($\mathbf{U},\mathbf{V}\in\bbU$) we obtain: 
\begin{align*}
\PKK{\lf}{\mf}(\omega;\fC) & =\pk{\mf}(\unn{\lf}{\mf}{\omega})=\unn{\lf}{\mf}{\omega}=\\
 & =\mathbf{V}\left(\arc{\mathbf{U}}(\omega)\right)=\mathbf{V}\left(\arc{\mathbf{U}}\left(\pk{\lf}(\omega)\right)\right)=\widetilde{Q}_{\mf,\lf}\left(\pk{\lf}(\omega)\right).
\end{align*}
 It is not hard to verify, that the system of mappings $\left(\widetilde{Q}_{\mf,\lf}\right)_{\lf,\mf\in\LkC}$
satisfies the conditions (\ref{eq:GPKconditions}). Therefore, by
the Definition \ref{Def:PeretvKoord} (item \ref{enu:GlPeretvKoord}),
we see, that $\left(\widetilde{Q}_{\mf,\lf}\right)_{\lf,\mf\in\LkC}$
is universal coordinate transform for $\fC$.      \EndProof  

Let $\left(\Ha,\left\Vert \cdot\right\Vert ,\left\langle \cdot,\cdot\right\rangle \right)$
be a Hilbert space over the real field and $\LHa$ be the space of
(homogeneous) linear continuous operators over the space $\Ha$. Denote
by $\LHax$ the space of all linear continuous operators over the
space $\Ha$, including non-homogeneous operators, that is $\LHax=\left\{ \mathbf{A}_{[\ba]}\,|\,\mathbf{\, A}\in\LHa,\,\ba\in\Ha\right\} $,
where $\mathbf{A}_{[\ba]}x=\mathbf{A}x+\ba$, $x\in\Ha$. The Hilbert
space $\Ha$ generates the coordinate space $\Haa=\left(\Ha,\tpT_{\Ha},\lsL_{\Ha},\rho_{\Ha},\left\Vert \cdot\right\Vert ,\left\langle \cdot,\cdot\right\rangle \right)$,
where $\rho_{\Ha}$ and $\tpT_{\Ha}$ are metrics and topology, generated
by the norm $\left\Vert \cdot\right\Vert $ on the space $\Ha$, as
well as $\lsL_{\Ha}$ is the natural linear structure of the space
$\Ha$. The \textbf{\emph{Minkowski space}} over the Hilbert space
$\Ha$ is defined as the Hilbert space $\MHa=\R\times\Ha=\left\{ \left(t,x\right)\,|\: t\in\R,\: x\in\Ha\right\} $,
equipped by the inner product and norm: $\left\langle \w_{1},\w_{2}\right\rangle =\left\langle \w_{1},\w_{2}\right\rangle _{\MHa}=t_{1}t_{2}+\left\langle x_{1},x_{2}\right\rangle $,
$\left\Vert \w_{1}\right\Vert =\left\Vert \w_{1}\right\Vert _{\MHa}=\left(t_{1}^{2}+\left\Vert x_{1}\right\Vert ^{2}\right)^{1/2}$
($\w_{i}=\left(t_{i},x_{i}\right)\in\MHa,$ $i\in\left\{ 1,2\right\} $)
(\cite{MyTmmTaxion01,MyTmmTaxion02}). In the space $\MHa$ we select
the next subspaces: 
\[
\Ha_{0}:=\left\{ \left(t,\oo\right)\,|\, t\in\R\right\} ,\quad\Ha_{1}:=\left\{ \left(0,x\right)\,|\, x\in\Ha\right\} ,
\]
with $\oo$ being zero vector. Then, $\MHa=\Ha_{0}\oplus\Ha_{1},$
where $\oplus$ means the orthogonal sum of subspaces. Denote: $\e:=\left(1,\oo\right)\in\MHa$.
Now, we introduce the orthogonal projectors on the subspaces $\Ha_{0}$
and $\Ha_{1}$: 
\begin{align}
\Xx\w & =\left(0,x\right)\in\Ha_{1};\quad\Tt\w=\left(t,\oo\right)=\Ttt\left(\w\right)\e\in\Ha_{0},\label{eq:TXdef}\\
 & \hspace{35mm}\text{where}\;\Ttt\left(\w\right)=t\quad\left(\w=\left(t,x\right)\in\MHa\right).\nonumber 
\end{align}

Any vector $V\in\Ha_{1}$ generates the following subspaces in the
space $\Ha_{1}$. 
\begin{gather*}
\Hao V=\sn\left\{ V\right\} ;\quad\Haoo V=\Ha_{1}\ominus\Hao V=\left\{ x\in\Ha_{1}\,|\,\left\langle x,V\right\rangle =0\right\} ,
\end{gather*}
where $\sn M$ denotes the linear span of the set $M\subseteq\MHa$.
We will denote by $\Xo V$ and $\Xoo V$ the orthogonal projectors
onto the subspaces $\Hao V$ and $\Haoo V$: 
\begin{gather}
\Xo V\w=\begin{cases}
\left\langle V,\w\right\rangle \left\Vert V\right\Vert ^{-2}V, & V\neq\oo\\
\oo, & V=\oo
\end{cases},\;\w\in\MHa;\quad\Xoo V=\Xx-\Xo V.\label{eq:XoOoDef}
\end{gather}
 Then for any vector $V\in\Ha_{1}$ we obtain the equality: 
\begin{equation}
\Tt+\Xx=\Tt+\Xo V+\Xoo V=\I_{\MHa},\label{eq:Tt+Xx}
\end{equation}
where $\I_{\MHa}$ is the identity operator on $\MHa$. 

Denote via $\PkH$ the set of all operators $\mathbf{S}\in\LMHax$,
which has the continuous inverse operator $\mathbf{S}^{-1}\in\LMHax$.
Operators $\mathbf{S}\in\PkH$ will be named as \textbf{\emph{coordinate
transform operators}}. Let, $\BBB$ be any base changeable set such,
that $\BsB\subseteq\Ha=\Zk\left(\Haa\right)$ and $\TMB=\left(\R,\leq\right)$,
where $\leq$ is the standard order in the field of real numbers $\R$.
Then~ $\BSB\subseteq\R\times\Ha=\MHa$. Any set  $\bbS\subseteq\PkH$
is transforming set of bijections relatively the $\BBB$ on $\Ha=\Zk\left(\Haa\right)$.
Therefore, we can put: 
\[
\Kim{\bbS,\BBB;\,\Ha}:=\Kim{\bbS,\BBB,\Haa}.
\]
 Now, we deduce the following corollary from Theorem \ref{Thm:KimUniversal}. 

\BeginNasl  \label{Nasl:KimUniversal:1} 

The kinematic set $\Kim{\bbS,\BBB;\,\Ha}$ allows universal coordinate
transform. 

\EndNasl 

Recall, that an operator $U\in\LHa$ is referred to as \textbf{\emph{unitary}}
on $\Ha$ if and only if $\exists\, U^{-1}\in\LHa$ and $\forall\, x\in\Ha\:\left\Vert Ux\right\Vert =\left\Vert x\right\Vert $.
Denote: 
\begin{gather*}
\UHa=\left\{ U\in\LHao\,|\: U\:\textrm{ is unitary on }\Ha_{1}\right\} ;\quad\BoHa=\left\{ x\in\Ha_{1}\,|\:\left\Vert x\right\Vert =1\right\} .
\end{gather*}

Consider any fixed values $c\in(0,\infty]$, $\lambda\in[0,\infty]\setminus\{c\}$,
$s\in\{-1,1\}$, $J\in\UHa$, $\bn\in\BoHa$, and $\ba\in\MHa$. For
an arbitrary vector $\w\in\MHa$ we put:
\begin{align}
\WlsnJ\w & :=\begin{cases}
\frac{\left(s\Ttt\left(\w\right)-\frac{\lambda}{c^{2}}\left\langle \bn,\w\right\rangle \right)}{\smxc{\lambda}}\e+J\left(\frac{\lambda\Ttt\left(\w\right)-s\left\langle \bn,\w\right\rangle }{\smxc{\lambda}}\bn+\Xoo{\bn}\w\right), & \lambda<\infty,\: c<\infty;\\
-\frac{\left\langle \bn,\w\right\rangle }{c}\e+J\left(c\Ttt\left(\w\right)\bn+\Xoo{\bn}\w\right), & \lambda=\infty,\: c<\infty;\\
s\Ttt(\w)\e+J\left(\left(\lambda\Ttt(\w)-s\left\langle \bn,\w\right\rangle \right)\bn+\Xoo{\bn}\w\right), & \lambda<\infty,\: c=\infty.
\end{cases}\nonumber \\
\WlsnJa & \w:=\WlsnJ(\w+\ba).\label{eq:WsnJdef}
\end{align}

In the case $c<\infty$ the operators of kind $\WlsnJ$ are generalized
Lorentz transforms, introduced in \cite{MyTmmTaxion01} (or in the
papers \cite{Recami1,Ricardo1,Hill_Cox}, for the case $\dim(\Ha)=3$).
Under the additional conditions $\lambda<c<\infty$, $\dim\left(\Ha\right)=3$,
$s=1$ the formula (\ref{eq:WsnJdef}) is equivalent to the formula
(28b) from \cite[page 43]{Myoler01}. That is why, in this case we
obtain the classical Lorentz transforms for inertial reference frame
in the most general form (with arbitrary orientation of axes). Moreover,
in the case $\dim\left(\Ha\right)=3$, $c<\infty$ the class of operators
$\LGpH=\left\{ \WlsnJ|\; s=1,\;\lambda<c\right\} $ coincides with
the full Lorentz group in the sense of \cite{Naimark1} (for more
details see \cite{MyTmm06}). The operators of kind $\WsnJ{\lambda,\infty}$~
($\lambda<\infty$) are Galilean transforms (it is not difficult prove,
that $\WsnJ{\lambda,\infty}=\lim_{c\to\infty}\WlsnJ$, where the convergence
is understood in the sense of uniform operator topology). 

\BeginAs  

For any $c\in(0,\infty]$, $\lambda\in[0,\infty]\setminus\{c\}$,
$s\in\{-1,1\}$, $J\in\UHa$, $\bn\in\BoHa$, and $\ba\in\MHa$ it
is true, that: 
\[
\WlsnJa\in\PkH.
\]

\EndAs  

\BeginProof 

Obviously it is sufficient to prove, that for $c\in(0,\infty]$, $\lambda\in[0,\infty]\setminus\{c\}$,
$s\in\{-1,1\}$, $J\in\UHa$ and $\bn\in\BoHa$ \textbf{\emph{the
operator $\WlsnJ$ has the continuous inverse $\WlsnJ^{-1}\in\LMHa$}}.
For the case $c<\infty$ the highlighted statement had been proved
in the paper \cite{MyTmmTaxion01}. 

That is why it remains to consider only the case $c=\infty$. Consider
any $\lambda\in[0,\infty)$, $s\in\{-1,1\}$, $J\in\UHa$ and $\bn\in\BoHa$.
It is easy to verify, that $\WsnJ{\lambda,\infty}\in\LMHa$. Now,
we are going to prove the equality: 
\begin{equation}
\WsnJ{\lambda,\infty}\Www{\lambda,\infty}{s,J\bn,J^{-1}}=\I_{\MHa}.\label{eq:WWI}
\end{equation}
 Chose any $\w\in\MHa$. According to (\ref{eq:WsnJdef}) we have:
\begin{align}
\WsnJ{\lambda,\infty} & \Www{\lambda,\infty}{s,J\bn,J^{-1}}\w=\WsnJ{\lambda,\infty}\wx,\quad\text{where}\label{eq:Wwxv}\\
\wx & =\Www{\lambda,\infty}{s,J\bn,J^{-1}}\w=s\Ttt(\w)\e+\nonumber \\
 & \qquad+J^{-1}\left(\left(\lambda\Ttt(\w)-s\left\langle J\bn,\w\right\rangle \right)J\bn+\Xoo{J\bn}\w\right).\nonumber 
\end{align}
 Next, applying (\ref{eq:TXdef}), (\ref{eq:XoOoDef}) and taking
into account that $J$ is the unitary operator on the subspace $\Ha_{1}\subseteq\MHa$,
we obtain: 
\begin{align*}
\Ttt(\wx) & =s\Ttt(\w);\\
\left\langle \bn,\wx\right\rangle  & =\left(\lambda\Ttt(\w)-s\left\langle J\bn,\w\right\rangle \right)\left\langle \bn,\bn\right\rangle +\left\langle \bn,J^{-1}\Xoo{J\bn}\w\right\rangle =\\
 & =\left(\lambda\Ttt(\w)-s\left\langle J\bn,\w\right\rangle \right)+\left\langle J\bn,\Xoo{J\bn}\w\right\rangle =\left(\lambda\Ttt(\w)-s\left\langle J\bn,\w\right\rangle \right);\\
\Xoo{\bn}\wx & =\left(\Xx-\Xo{\bn}\right)\wx=\\
 & =J^{-1}\left(\left(\lambda\Ttt(\w)-s\left\langle J\bn,\w\right\rangle \right)J\bn+\Xoo{J\bn}\w\right)-\left\langle \bn,\wx\right\rangle \bn=\\
 & =J^{-1}\left(\left\langle \bn,\wx\right\rangle J\bn+\Xoo{J\bn}\w\right)-\left\langle \bn,\wx\right\rangle \bn=J^{-1}\Xoo{J\bn}\w.
\end{align*}
 Herefrom, using (\ref{eq:Wwxv}), (\ref{eq:WsnJdef}) and (\ref{eq:Tt+Xx})
we deduce: 
\begin{align*}
 & \hspace{-1cm}\WsnJ{\lambda,\infty}\Www{\lambda,\infty}{s,J\bn,J^{-1}}\w=\\
 & =s\Ttt(\wx)\e+J\left(\left(\lambda\Ttt(\wx)-s\left\langle \bn,\wx\right\rangle \right)\bn+\Xoo{\bn}\wx\right)=s\left(s\Ttt(\w)\right)\e+\\
 & \qquad+J\left(\left(\lambda\left(s\Ttt(\w)\right)-s\left(\lambda\Ttt(\w)-s\left\langle J\bn,\w\right\rangle \right)\right)\bn+J^{-1}\Xoo{J\bn}\w\right)=\\
 & =\Ttt(\w)\e+\left\langle J\bn,\w\right\rangle J\bn+\Xoo{J\bn}\w=\Tt\w+\Xo{J\bn}\w+\Xoo{J\bn}\w=\w.
\end{align*}
 Equality (\ref{eq:WWI}) is proved. Applying the equality (\ref{eq:WWI})
to the operator $\Www{\lambda,\infty}{s,J\bn,J^{-1}}$, we obtain
the equality: $\Www{\lambda,\infty}{s,J\bn,J^{-1}}\WsnJ{\lambda,\infty}=\I_{\MHa}$.
Thus, $\WlsnJ^{-1}=\Www{\lambda,\infty}{s,J\bn,J^{-1}}\in\LMHa$.
   \EndProof 

For $0<c\leq\infty$ we introduce the following classes of linear
(non-homogeneous) operators: 
\begin{align*}
\PTH & :=\left\{ \WlsnJa\,|\; s\in\left\{ -1,1\right\} ,\:\lambda\in[0,\infty]\setminus\{c\},\right.\\
 & \left.\qquad\qquad\qquad\qquad\quad\;\bn\in\BoHa,\: J\in\UHa,\:\ba\in\MHa\right\} ;\\
\PTpH & :=\left\{ \WlsnJa\in\PTH\,|\; s=1\right\} ;\\
\PGH & :=\left\{ \WlsnJa\in\PTH\,|\;\lambda<c\right\} ;\\
\PGpH & :=\left\{ \WlsnJa\in\PGH\,|\: s=1\right\} 
\end{align*}
(It is apparently, that $\PT{\Ha,\infty}=\PG{\Ha,\infty}$, $\PTp{\Ha,\infty}=\PGp{\Ha,\infty}$).~
Using the introduced classes of operators, we may define the following
kinematic sets:  
\begin{align*}
\KPTn{\Ha,\BBB,c} & :=\Kim{\PTH,\BBB;\,\Ha};\\
\KPT{\Ha,\BBB,c} & :=\Kim{\PTpH,\BBB;\,\Ha};\\
\KPn{\Ha,\BBB,c} & :=\Kim{\PGH,\BBB;\,\Ha};\\
\KP{\Ha,\BBB,c} & :=\Kim{\PGpH,\BBB;\,\Ha}.
\end{align*}

In the case $\dim(\Ha)=3$, $c<\infty$ the kinematic set $\KP{\Ha,\BBB,c}$
represents the simplest mathematically strict model of the kinematics
of special relativity theory in inertial frames of reference. Kinematic
set $\KPn{\Ha,\BBB,c}$ is constructed on the basis of general Lorentz-Poincare
group, and it includes apart from usual reference frames (with positive
direction of time), which have understandable physical interpretation,
also reference frames with negative direction of time. Kinematic
sets $\KPT{\Ha,\BBB,c}$ and $\KPTn{\Ha,\BBB,c}$ include apart from
standard (``tardyon'') reference frames also ``tachyon'' reference
frames,  which are moving relatively the ``tardyon'' reference
frames with velocity, greater than the velocity of light $c$. Kinematic
set $\KP{\Ha,\BBB,\infty}=\KPT{\Ha,\BBB,\infty}$ in the case $\dim(\Ha)=3$,
$c=\infty$ represents the mathematically strict model of the Galilean
kinematics in the inertial frames of reference. ~ The next corollary
follows from Theorem \ref{Thm:KimUniversal}. 

\BeginNasl  

Kinematic sets $\KPTn{\Ha,\BBB,c}$, $\KPT{\Ha,\BBB,c}$, $\KPn{\Ha,\BBB,c}$,
$\KP{\Ha,\BBB,c}$ allow universal coordinate transform. 

\EndNasl  

\BeginRmk 

From the results of the works \cite{MyTmmTaxion01,MyTmm06} it follows,
that the sets of operators $\PGH$ and $\PGpH$ form the groups of
operators over the space $\MHa$ (in particular case $\dim(\Ha)=3$
the group of operators $\PGpH$ coincides with the classical Poincare
group in four-dimensional Minkowski space-time). At the same time,
in the \cite{MyTmmTaxion02} it is proved, that the classes of operators
$\PTH$ and $\PTpH$ do not form a group over $\MHa$. This means,
that the kinematics $\KPTn{\Ha,\BBB,c}$ and $\KPT{\Ha,\BBB,c}$,
constructed on the basis of these classes, do not satisfy the relativity
principle, because, according to Theorem \ref{Thm:KimUniversal},
the subset of coordinate transforms (\ref{eq:Kim:UniversalPk}), providing
transition from one reference frame to all other, is different for
different frames. But, in kinematics $\KPTn{\Ha,\BBB,c}$ and $\KPT{\Ha,\BBB,c}$
the relativity principle violated only in the superluminal diapason,
because the kinematics sets $\KPTn{\Ha,\BBB,c}$ and $\KPT{\Ha,\BBB,c}$
are formed by the ``addition'' of new, superlight reference frames
to the kinematics sets $\KPn{\Ha,\BBB,c}$ and $\KP{\Ha,\BBB,c}$,
which satisfy the principle of relativity. It should be noted that
the principle of relativity is only one of experimentally established
facts. Therefore, it is possible that this principle is not satisfied
when we exit out of the light barrier.  Possibility of revision of
the relativity principle is now discussed in the physical literature
(see for example, \cite{BaccettiTateVisser,BaccettiTateVisser02,BaccettiTateVisser03,EoloDiCasola01,S_Liberati01,Gao_shan01,Kholmetskii01}). 

\EndRmk

\section{Kinematic Sets, which do not Allow Universal Coordinate Transform. }

In this section, it is constructed one interesting class of kinematic
sets, in which every particle at each time moment can can have its
own ``velocity of light''. On a physical level, the similar models
(with particle-dependent velocity of light) were considered in the
papers \cite{Glashow01,Glashow02,Drago01,Gertov01,Kazarian01}. 

Let a set $\Vf\subseteq(0,\infty]$ be such, that $\Vf\neq\emptyset$
and $(0,\infty]\setminus\Vf\neq\emptyset$. Denote: 
\begin{gather*}
\Ha_{\Vf}:=\Ha\times\Vf=\left\{ \left(x,c\right)\,|\: x\in\Ha,\: c\in\Vf\right\} ;\quad\MHac:=\R\times\Ha_{\Vf}.
\end{gather*}
 The set $\MHac$ will be named as the Minkowski space \textbf{\emph{with
the set of forbidden velocities}} $\Vf$ over $\Ha$. The set $\Cvp:=[0,\infty]\setminus\Vf$
will be named as the \textbf{\emph{set of allowed velocities}} for
the space $\MHac$. 

For an arbitrary $\omega=\left(t,\left(x,c\right)\right)\in\MHac$
we put $\omega^{*}:=\left(t,x\right)\in\MHa.$ Also for $\lambda\in\Cvp$,
$s\in\{-1,1\}$, $J\in\UHa$, $\bn\in\BoHa$, $\ba\in\MHa$ and $\omega=\left(t,\left(x,c\right)\right)\in\MHac$
we introduce the denotation: 
\begin{equation}
\WLsnJa\omega:=\left(\tm{\WlsnJa\omega^{*}},\:\left(\bs{\WlsnJa\omega^{*}},c\right)\right).\label{eq:WlVf[snJa]def}
\end{equation}
 Therefore, for any $\omega=\left(t,\left(x,c\right)\right)\in\MHac$
we have the equality: 
\begin{equation}
\left(\WLsnJa\omega\right)^{*}=\WlsnJa\omega^{*}.\label{eq:Wlc[snJa]w*}
\end{equation}

\BeginAs 

For arbitrary $\lambda\in\Cvp$, $s\in\{-1,1\}$, $J\in\UHa$, $\bn\in\BoHa$,
$\ba\in\MHa$ the mapping $\WLsnJa$ is bijection on $\MHac$. 

\EndAs  

\BeginProof 

Suppose, that $\WLsnJa\omega_{1}=\WLsnJa\omega_{2}$, where $\omega_{1}=\left(t_{1},\left(x_{1},c_{1}\right)\right)\in\MHac$,
$\omega_{2}=\left(t_{2},\left(x_{2},c_{2}\right)\right)\in\MHac$.
Then, 
\begin{gather*}
\left(\tm{\WsnJa{\lambda,c_{1}}\omega_{1}^{*}},\:\left(\bs{\WsnJa{\lambda,c_{1}}\omega_{1}^{*}},c_{1}\right)\right)=\qquad\\
\qquad=\left(\tm{\WsnJa{\lambda,c_{2}}\omega_{2}^{*}},\:\left(\bs{\WsnJa{\lambda,c_{2}}\omega_{2}^{*}},c_{2}\right)\right).
\end{gather*}
 Consequently, $c_{1}=c_{2}$. Hence, we have proved the equalities:
\begin{align*}
\tm{\WsnJa{\lambda,c_{1}}\omega_{1}^{*}} & =\tm{\WsnJa{\lambda,c_{1}}\omega_{2}^{*}}\\
\bs{\WsnJa{\lambda,c_{1}}\omega_{1}^{*}} & =\bs{\WsnJa{\lambda,c_{1}}\omega_{2}^{*}}.
\end{align*}
 Therefore, $\WsnJa{\lambda,c_{1}}\omega_{1}^{*}=\WsnJa{\lambda,c_{1}}\omega_{2}^{*}$.
And, taking into account the fact, that the mapping $\WsnJa{\lambda,c_{1}}$
is bijection on $\MHa$, we conclude, that, $\omega_{1}^{*}=\omega_{2}^{*}$,
ie $t_{1}=t_{2}$, $x_{1}=x_{2}$. Hence, $\omega_{1}=\left(t_{1},\left(x_{1},c_{1}\right)\right)=\left(t_{2},\left(x_{2},c_{2}\right)\right)=\omega_{2}$.
Thus, the mapping $\WsnJa{\lambda;\Vf}$ is one-to-one correspondence. 

Now it remains to prove, that $\WsnJa{\lambda;\Vf}$ reflects the
set $\MHac$ on $\MHac$. Consider any $\omega=\left(t,\left(x,c\right)\right)\in\MHac$.
Denote: 
\[
\widetilde{\omega}:=\left(\tm{\arc{\left(\WlsnJa\right)}\omega^{*}},\,\left(\bs{\arc{\left(\WlsnJa\right)}\omega^{*}},c\right)\right).
\]
 Then, 
\begin{align*}
\widetilde{\omega}^{*} & =\left(\tm{\arc{\left(\WlsnJa\right)}\omega^{*}},\,\bs{\arc{\left(\WlsnJa\right)}\omega^{*}}\right)=\\
 & =\arc{\left(\WlsnJa\right)}\omega^{*}.
\end{align*}
 Consequently, $\WlsnJa\widetilde{\omega}^{*}=\omega^{*}$. Hence,
\begin{align*}
\WLsnJa\widetilde{\omega} & =\left(\tm{\WlsnJa\widetilde{\omega}^{*}},\:\left(\bs{\WlsnJa\widetilde{\omega}^{*}},c\right)\right)=\\
 & =\left(\tm{\omega^{*}},\:\left(\bs{\omega^{*}},c\right)\right)=\left(t,\left(x,c\right)\right)=\omega.
\end{align*}
Thus, $\WLsnJa$ is bijection from $\MHac$ onto $\MHac$.    
\EndProof 

Denote: 
\begin{align*}
\PTHC & :=\left\{ \WLsnJa\,|\:\lambda\in\Cvp,\: s\in\{-1,1\},\right.\\
 & \left.\phantom{\Cvp|}\qquad\qquad\qquad\qquad\;\: J\in\UHa,\:\bn\in\BoHa,\:\ba\in\MHa\right\} ;\\
\PTpHC & :=\left\{ \WLsnJa\in\PTHC\,|\: s=1\right\} .
\end{align*}

Let, $\BBB$ be a base changeable set such, that $\BsB\subseteq\Ha_{\Vf}$,
$\TMB=\left(\R,\leq\right)$. Then we have, $\BSB\subseteq\R\times\Ha_{\Vf}=\MHac$.
Hence, we deliver the following kinematic multi-projectors: 
\begin{align}
\PTHC^{\wedge} & =\left(\left(\left(\R,\leq\right),\Ha_{\Vf},\mathbf{S},\Haa,\mathbf{q}\right)|\:\mathbf{S}\in\PTHC\right);\nonumber \\
\PTpHC^{\wedge} & =\left(\left(\left(\R,\leq\right),\Ha_{\Vf},\mathbf{S},\Haa,\mathbf{q}\right)|\:\mathbf{S}\in\PTpHC\right),\quad\text{where}\nonumber \\
\mathbf{q}(\xxv) & =x\quad\left(\forall\:\xxv=\left(x,c\right)\in\Ha_{\Vf}\right)\label{eq:def:PTmprj+q}
\end{align}
 for $\BBB$. In accordance with Theorem \ref{Thm:MultiImg(KM)},
we can denote: 
\begin{align*}
\KPTn{\Ha,\BBB;\Vf} & :=\Kimm{\PTHC^{\wedge},\BBB};\\
\KPT{\Ha,\BBB;\Vf} & :=\Kimm{\PTpHC^{\wedge},\BBB}.
\end{align*}

It turns out, that the kinematic sets $\KPTn{\Ha,\BBB;\Vf}$ and $\KPT{\Ha,\BBB;\Vf}$,
in the general case, do not allow universal coordinate transform.
More precisely, they allow universal coordinate transform if and only
if only one value of forbidden velocity $c\in(0,\infty]$ is actually
realized. In the last case, kinematics in $\KPTn{\Ha,\BBB;\Vf}$ or
$\KPT{\Ha,\BBB;\Vf}$ can be reduced to kinematics of type $\KPTn{\Ha,\BBB,c}$
or $\KPT{\Ha,\BBB,c}$ (for $c<\infty$), and to Galilean kinematics
(for $c=\infty$). 

\BeginThm \label{Thm:KPT(HBVf)_universPK}

Let the set of forbidden velocities $\Vf$ be separated from zero
(ie there exists a number $\eta>0$ such, that $\Vf\subseteq[\eta,\infty]$). 

Kinematic set $\KPT{\Ha,\BBB;\Vf}$ allows universal coordinate transform
if and only if there don't exist elementary states $\xxv_{1}=\left(x_{1},c_{1}\right),\:\xxv_{2}=\left(x_{2},c_{2}\right)\in\BsB$
such, that $c_{1}\neq c_{2}$. 

\EndThm 

To prove Theorem \ref{Thm:KPT(HBVf)_universPK} we need the following
two lemmas. 

\BeginLem \label{Lem:(Wlc1[snj]w1=00003DWlc2[snj]w2)}~ Chose any
fixed $c_{1},c_{2}\in(0,\infty]$, $c_{1}\neq c_{2}$, $s\in\{-1,1\}$
and $J\in\UHa$.

Then, for any number $\varepsilon\in\left(0,\min\left(c_{1},c_{2}\right)\right)$,
and arbitrary vectors $\w_{1},\w_{2}\in\MHa$ such, that $\w_{1}\neq\w_{2}$
there exist $\lambda\in(0,\varepsilon)$, $\bn\in\BoHa$ and $\ba\in\MHa$,
for which the following equality holds: 
\[
\Www{\lambda,c_{1}}{s,\bn,J;\ba}\w_{1}=\Www{\lambda,c_{2}}{s,\bn,J;\ba}\w_{2}.
\]

\EndLem 

\BeginProof 

Further, for convenience, we assume, that $c_{1}<c_{2}$. Obviously,
this assumption does not restrict the the generality of our conclusions. 

\textbf{1.} At first, we are going to prove Lemma in the special case
$\w_{1}=\oo$, $\w_{2}=\w\neq\oo$. Consider any, $\varepsilon\in\left(0,\min\left(c_{1},c_{2}\right)\right)$.
According to the specifics of this case, we should find $\lambda\in(0,\varepsilon)$,
$\bn\in\BoHa$ and $\ba\in\MHa$, such, that:
\begin{equation}
\Www{\lambda,c_{1}}{s,\bn,J;\ba}\oo=\Www{\lambda,c_{2}}{s,\bn,J;\ba}\w.\label{eq:condit(lmbd,n,a)0}
\end{equation}
Taking into account (\ref{eq:WsnJdef}), we can rewrite the last condition
in the form: 
\begin{equation}
\Www{\lambda,c_{1}}{s,\bn,J}\ba=\Www{\lambda,c_{2}}{s,\bn,J}\left(\w+\ba\right).\label{eq:condit(lmbd,n,a)1}
\end{equation}

Denote: 
\begin{equation}
t:=\Ttt\left(\w\right),\qquad x:=\Xx\w.\label{eq:txDef}
\end{equation}
Then we can write, $\w=t\e+x.$ 

Consider any fixed vector $\bn_{0}\in\BoHa$. Denote: 
\begin{equation}
\bn:=\begin{cases}
\frac{x}{\left\Vert x\right\Vert }, & x\neq\oo\\
\bn_{0}, & x=\oo.
\end{cases}\label{eq:Vect_n_def}
\end{equation}
 Then, we have:
\begin{gather}
\begin{array}{rl}
x\:= & \left\Vert x\right\Vert \bn,\\
\left\langle \bn,\w\right\rangle \:= & \left\langle \bn,x\right\rangle =\left\Vert x\right\Vert ,\\
\Xoo{\bn}\w\:= & \Xx\w-\left\langle \bn,\w\right\rangle \bn=x-\left\Vert x\right\Vert \bn=x-x=\oo.
\end{array}\label{eq:(n,w);Xoo[n]w}
\end{gather}

Vector $\ba$ we seek in the form: 
\begin{equation}
\ba=\tau\e+\mu\bn,\qquad\text{where}\quad\tau,\mu\in\R.\label{eq:vect_a_search_repr}
\end{equation}

\textbf{1.a)} At first we consider the case $c_{1},c_{2}<\infty$. 

Substituting the value of the vector $\ba$ from (\ref{eq:vect_a_search_repr})
into the condition (\ref{eq:condit(lmbd,n,a)1}) and applying (\ref{eq:txDef}),
(\ref{eq:(n,w);Xoo[n]w}), (\ref{eq:WsnJdef}), we obtain the following
condition: 
\begin{align}
 & \hspace{-15mm}\left(s\tau-\frac{\lambda}{c_{1}^{2}}\mu\right)\gamma\left(\frac{\lambda}{c_{1}}\right)\e+\left(\lambda\tau-s\mu\right)\gamma\left(\frac{\lambda}{c_{1}}\right)\, J\bn=\nonumber \\
 & =\left(s\left(t+\tau\right)-\frac{\lambda}{c_{2}^{2}}\left(\left\Vert x\right\Vert +\mu\right)\right)\gamma\left(\frac{\lambda}{c_{2}}\right)\e+\nonumber \\
 & \qquad+\left(\lambda\left(t+\tau\right)-s\left(\left\Vert x\right\Vert +\mu\right)\right)\gamma\left(\frac{\lambda}{c_{2}}\right)\, J\bn,\nonumber \\
 & \qquad\qquad\qquad\text{where}\quad\gamma(\xi)=\frac{1}{\sqrt{\left|1-\xi^{2}\right|}},\quad\xi\geq0,\:\xi\neq1.\label{eq:func_gam_def}
\end{align}
 Taking into account orthogonality of the vector $\e$ to the subspace
$\Ha_{1}$ and unitarity of the operator $J$ on the subspace $\Ha_{1}$,
we get the following system of equations: 
\begin{gather}
\begin{cases}
\left(s\tau-\frac{\lambda}{c_{1}^{2}}\mu\right)\gamma\left(\frac{\lambda}{c_{1}}\right)=\left(s\left(t+\tau\right)-\frac{\lambda}{c_{2}^{2}}\left(\left\Vert x\right\Vert +\mu\right)\right)\gamma\left(\frac{\lambda}{c_{2}}\right)_{\phantom{1_{1_{1}}}}\\
\left(\lambda\tau-s\mu\right)\gamma\left(\frac{\lambda}{c_{1}}\right)=\left(\lambda\left(t+\tau\right)-s\left(\left\Vert x\right\Vert +\mu\right)\right)\gamma\left(\frac{\lambda}{c_{2}}\right)
\end{cases}\label{eq:SysEq(l,t,m)1}
\end{gather}
By means of simple transformations, the system (\ref{eq:SysEq(l,t,m)1})
can be reduced to the form: 
\begin{gather}
\begin{cases}
\tau\left(\gamma\left(\frac{\lambda}{c_{2}}\right)-\gamma\left(\frac{\lambda}{c_{1}}\right)\right)=\lambda s\left(\left(\frac{\left\Vert x\right\Vert +\mu}{c_{2}^{2}}\right)\gamma\left(\frac{\lambda}{c_{2}}\right)-\frac{\mu}{c_{1}^{2}}\gamma\left(\frac{\lambda}{c_{1}}\right)\right)-t\gamma\left(\frac{\lambda}{c_{2}}\right)_{\phantom{1_{1_{1}}}}\\
\lambda\tau\left(\gamma\left(\frac{\lambda}{c_{2}}\right)-\gamma\left(\frac{\lambda}{c_{1}}\right)\right)=s\left(\left(\left\Vert x\right\Vert +\mu\right)\gamma\left(\frac{\lambda}{c_{2}}\right)-\mu\gamma\left(\frac{\lambda}{c_{1}}\right)\right)-\lambda t\gamma\left(\frac{\lambda}{c_{2}}\right)
\end{cases}\label{eq:SysEq(l,t,m)2}
\end{gather}
 Replacing the expression $\tau\left(\gamma\left(\frac{\lambda}{c_{2}}\right)-\gamma\left(\frac{\lambda}{c_{1}}\right)\right)$
in the second equation of the system (\ref{eq:SysEq(l,t,m)2}) by
the right-hand side of the first equation of this system, we deliver
the equation: 
\[
\lambda^{2}\left(\frac{\left\Vert x\right\Vert +\mu}{c_{2}^{2}}\gamma\left(\frac{\lambda}{c_{2}}\right)-\frac{\mu}{c_{1}^{2}}\gamma\left(\frac{\lambda}{c_{1}}\right)\right)=\left(\left\Vert x\right\Vert +\mu\right)\gamma\left(\frac{\lambda}{c_{2}}\right)-\mu\gamma\left(\frac{\lambda}{c_{1}}\right).
\]
After a simple transformations the last equation takes a form: 
\begin{equation}
\frac{\left(\left\Vert x\right\Vert +\mu\right)\left(1-\frac{\lambda^{2}}{c_{2}^{2}}\right)}{\Smxc{\lambda}{c_{2}}}-\frac{\mu\left(1-\frac{\lambda^{2}}{c_{1}^{2}}\right)}{\Smxc{\lambda}{c_{1}}}=0.\label{eq:Eq(l,m)1}
\end{equation}

Now, we introduce the denotations: 
\begin{equation}
\Phi_{1}(y):=\sign(y)\sqrt{\left|y\right|};\quad\Phi_{2}(y)=y\left|y\right|\qquad(y\in\R).\label{eq:Fi1(y)def}
\end{equation}
In the case $y\neq0$ the function $\Phi_{1}(y)$ may be represented
in the form, $\Phi_{1}(y)=\frac{y}{\sqrt{\left|y\right|}}$. 

In view of denotation (\ref{eq:Fi1(y)def}) the equation (\ref{eq:Eq(l,m)1})
becomes: 
\[
\Phi_{1}\left(\left(\left\Vert x\right\Vert +\mu\right)\left|\left\Vert x\right\Vert +\mu\right|\left(1-\frac{\lambda^{2}}{c_{2}^{2}}\right)\right)=\Phi_{1}\left(\mu\left|\mu\right|\left(1-\frac{\lambda^{2}}{c_{1}^{2}}\right)\right).
\]
Taking into account, that the function $\Phi_{1}$ is strictly monotone
on $\R$, we get the equation: 
\[
\left(\left\Vert x\right\Vert +\mu\right)\left|\left\Vert x\right\Vert +\mu\right|\left(1-\frac{\lambda^{2}}{c_{2}^{2}}\right)=\mu\left|\mu\right|\left(1-\frac{\lambda^{2}}{c_{1}^{2}}\right),
\]
which after a simple transformations is reduced to the form: 
\begin{align}
\lambda^{2}\left(\Phi_{2}\left(\frac{\left\Vert x\right\Vert +\mu}{c_{2}}\right)-\Phi_{2}\left(\frac{\mu}{c_{1}}\right)\right) & =\Phi_{2}\left(\left\Vert x\right\Vert +\mu\right)-\Phi_{2}\left(\mu\right).\label{eq:Eq(l,m)2}
\end{align}
 Since $c_{1}<c_{2}$, then for $\mu<-\left\Vert x\right\Vert $ we
have $\frac{\left\Vert x\right\Vert +\mu}{c_{2}}>\frac{\mu}{c_{1}}$.
Therefore, taking into account, that the function $\Phi_{2}$ is strictly
increasing on $\R$, we may define the function: 
\[
\Phi_{3;x}(\mu)=\sqrt{\frac{\Phi_{2}\left(\left\Vert x\right\Vert +\mu\right)-\Phi_{2}\left(\mu\right)}{\Phi_{2}\left(\frac{\left\Vert x\right\Vert +\mu}{c_{2}}\right)-\Phi_{2}\left(\frac{\mu}{c_{1}}\right)}}=\sqrt{\frac{\mu^{2}-\left(\left\Vert x\right\Vert +\mu\right)^{2}}{\left(\frac{\mu}{c_{1}}\right)^{2}-\left(\frac{\left\Vert x\right\Vert +\mu}{c_{2}}\right)^{2}}},\qquad\mu<-\left\Vert x\right\Vert .
\]
It is easy to verify, that $\Phi_{3;x}(\mu)\to0$, $\mu\to-\infty$.
Hence, there exists the number $\mu_{0}<-\left\Vert x\right\Vert $
such, that $\Phi_{3;x}\left(\mu_{0}\right)\in[0,\varepsilon)$. 

In the case $x\neq\oo$ we have $\Phi_{3;x}(\mu)>0$ for all $\mu$
such, that $\mu<-\left\Vert x\right\Vert $. In the case $x=\oo$,
the equation (\ref{eq:Eq(l,m)2}) becomes the true equality for $\mu=0$
and arbitrary $\lambda\in\R$.~ That is why, if we put: 
\begin{equation}
\mu:=\begin{cases}
\mu_{0}, & x\neq\oo\\
0, & x=\oo
\end{cases};\qquad\lambda:=\begin{cases}
\Phi_{3;x}\left(\mu_{0}\right), & x\neq\oo\\
\frac{\varepsilon}{2}, & x=\oo,
\end{cases}\label{eq:MuLmbd_Def}
\end{equation}
 we will obtain the values $\mu\in\R$ and $\lambda\in(0,\varepsilon)$,
for which the equality (\ref{eq:Eq(l,m)2}) holds.

 Since $0<\lambda<\varepsilon<\min\left(c_{1},c_{2}\right)$, then
for values $\lambda,\mu$, determined by the formula (\ref{eq:MuLmbd_Def}),
the second equation from the system (\ref{eq:SysEq(l,t,m)2}) takes
the form: 
\begin{equation}
\lambda\tau\left(\frac{1}{\Sxc{\lambda}{c_{2}}}-\frac{1}{\Sxc{\lambda}{c_{1}}}\right)=s\left(\frac{\left\Vert x\right\Vert +\mu}{\Sxc{\lambda}{c_{2}}}-\frac{\mu}{\Sxc{\lambda}{c_{1}}}\right)-\frac{\lambda t}{\Sxc{\lambda}{c_{2}}},\label{eq:Eq_tau1}
\end{equation}
 where, considering that $\lambda>0$ and $c_{1}<c_{2}$, we have
$\lambda\left(\frac{1}{\Sxc{\lambda}{c_{2}}}-\frac{1}{\Sxc{\lambda}{c_{1}}}\right)\neq0$.
Hence, the number $\tau$ is uniquely determined by the equality (\ref{eq:Eq_tau1}).
Then, the vector $\ba$ we calculate by the formula (\ref{eq:vect_a_search_repr}).
And, substituting the delivered values of the parameters $\lambda\in(0,\varepsilon)$,
$\bn\in\BoHa$ and $\ba\in\MHa$ into (\ref{eq:condit(lmbd,n,a)1}),
we guarantee the valid equality. In the case $c_{2},c_{2}<\infty$
and $\w_{1}=\oo$, Lemma is proved. 

\textbf{1.b)} Thus, it remains to consider only the case $c_{2}=\infty$,
$c_{1}<\infty$ ($\w_{1}=\oo$, $\w_{2}=\w\neq\oo$). Note, that the
case $c_{1}=\infty$ is impossible, because $c_{1}<c_{2}$. 

Substituting the value of the vector $\ba$ from (\ref{eq:vect_a_search_repr})
into the condition (\ref{eq:condit(lmbd,n,a)1}) and applying (\ref{eq:txDef}),
(\ref{eq:(n,w);Xoo[n]w}), (\ref{eq:WsnJdef}), we obtain the following
condition: 
\begin{align*}
\left(s\tau-\frac{\lambda}{c_{1}^{2}}\mu\right)\gamma\left(\frac{\lambda}{c_{1}}\right)\e & +\left(\lambda\tau-s\mu\right)\gamma\left(\frac{\lambda}{c_{1}}\right)\, J\bn=\\
 & =s\left(t+\tau\right)\e+\left(\lambda\left(t+\tau\right)-s\left(\left\Vert x\right\Vert +\mu\right)\right)\, J\bn.
\end{align*}
 Hence, taking into account orthogonality of the vector $\e$ to the
subspace $\Ha_{1}$ and unitarity of the operator $J$ on the subspace
$\Ha_{1}$, we get the following system of equations: 
\begin{gather}
\begin{cases}
\left(s\tau-\frac{\lambda}{c_{1}^{2}}\mu\right)\gamma\left(\frac{\lambda}{c_{1}}\right)=s\left(t+\tau\right)\\
\left(\lambda\tau-s\mu\right)\gamma\left(\frac{\lambda}{c_{1}}\right)=\lambda\left(t+\tau\right)-s\left(\left\Vert x\right\Vert +\mu\right)
\end{cases}\label{eq:8SysEq(l,t,m)1}
\end{gather}
 After a simple transformations, the system (\ref{eq:8SysEq(l,t,m)1})
may be reduced to the form: 
\begin{gather}
\begin{cases}
\tau\left(1-\gamma\left(\frac{\lambda}{c_{1}}\right)\right)=-\lambda s\,\frac{\mu}{c_{1}^{2}}\gamma\left(\frac{\lambda}{c_{1}}\right)-t\\
\lambda\tau\left(1-\gamma\left(\frac{\lambda}{c_{1}}\right)\right)=s\left(\left\Vert x\right\Vert +\mu-\mu\gamma\left(\frac{\lambda}{c_{1}}\right)\right)-\lambda t
\end{cases}\label{eq:8SysEq(l,t,m)2}
\end{gather}
 Replacing the expression $\tau\left(1-\gamma\left(\frac{\lambda}{c_{1}}\right)\right)$
in the second equation of the system (\ref{eq:8SysEq(l,t,m)2}) by
the right-hand side of the first equation of this system, we obtain
the equation: 
\[
-\lambda^{2}\frac{\mu}{c_{1}^{2}}\gamma\left(\frac{\lambda}{c_{1}}\right)=\left\Vert x\right\Vert +\mu-\mu\gamma\left(\frac{\lambda}{c_{1}}\right),
\]
which, by means of a simple transformations takes a form:

\begin{equation}
\Phi_{1}\left(\left(\left\Vert x\right\Vert +\mu\right)\left|\left\Vert x\right\Vert +\mu\right|\right)=\Phi_{1}\left(\mu\left|\mu\right|\left(1-\frac{\lambda^{2}}{c_{1}^{2}}\right)\right),\label{eq:8Eq(l,m)1}
\end{equation}
 where the function $\Phi_{1}$ is determined by the formula (\ref{eq:Fi1(y)def}).
Taking into account, that the function $\Phi_{1}$ is strictly monotone
on $\R$, we get the equation: 
\[
\left(\left\Vert x\right\Vert +\mu\right)\left|\left\Vert x\right\Vert +\mu\right|=\mu\left|\mu\right|\left(1-\frac{\lambda^{2}}{c_{1}^{2}}\right),
\]
which after a simple transformations is reduced to the form: 
\begin{align}
-\lambda^{2}\Phi_{2}\left(\frac{\mu}{c_{1}}\right) & =\Phi_{2}\left(\left\Vert x\right\Vert +\mu\right)-\Phi_{2}\left(\mu\right).\label{eq:8Eq(l,m)2}
\end{align}
 Therefore, taking into account, that the function $\Phi_{2}(y)=y\left|y\right|$
is strictly increasing on $\R$, we may define the function: 
\[
\Phi_{3;x}^{\infty}(\mu)=\sqrt{\frac{\Phi_{2}\left(\left\Vert x\right\Vert +\mu\right)-\Phi_{2}\left(\mu\right)}{-\Phi_{2}\left(\frac{\mu}{c_{1}}\right)}}=\sqrt{\frac{\mu^{2}-\left(\left\Vert x\right\Vert +\mu\right)^{2}}{\left(\frac{\mu}{c_{1}}\right)^{2}}},\qquad\mu<-\left\Vert x\right\Vert .
\]
It is easy to verify, that $\Phi_{3;x}^{\infty}(\mu)\to0$, $\mu\to-\infty$.
Hence, there exists the number $\mu_{0}<-\left\Vert x\right\Vert $
such, that $\Phi_{3;x}^{\infty}\left(\mu_{0}\right)\in[0,\varepsilon)$. 

In the case $x\neq\oo$ we have $\Phi_{3;x}^{\infty}(\mu)>0$ for
all $\mu$ such, that $\mu<-\left\Vert x\right\Vert $. In the case
$x=\oo$, the equation (\ref{eq:8Eq(l,m)2}) becomes the true equality
for $\mu=0$ and arbitrary $\lambda\in\R$. That is why, if we put:
\begin{equation}
\mu:=\begin{cases}
\mu_{0}, & x\neq\oo\\
0, & x=\oo
\end{cases}\qquad\lambda:=\begin{cases}
\Phi_{3;x}^{\infty}\left(\mu_{0}\right), & x\neq\oo\\
\frac{\varepsilon}{2}, & x=\oo,
\end{cases}\label{eq:MuLmbd_Def1}
\end{equation}
we will obtain the values $\mu\in\R$ and $\lambda\in(0,\varepsilon)$,
for which the equality (\ref{eq:8Eq(l,m)2}) is true. 

Since $0<\lambda<\varepsilon<\min\left(c_{1},c_{2}\right)=c_{1}$,
then for values $\lambda,\mu$, determined by the formula (\ref{eq:MuLmbd_Def1}),
the second equation from the system (\ref{eq:8SysEq(l,t,m)2}) may
be rewritten in the form: 
\begin{equation}
\lambda\tau\left(1-\frac{1}{\Sxc{\lambda}{c_{1}}}\right)=s\left(\left\Vert x\right\Vert +\mu-\frac{\mu}{\Sxc{\lambda}{c_{1}}}\right)-\lambda t,\label{eq:8Eq_tau1}
\end{equation}
 where, considering that $\lambda,c_{1}>0$, we have, $\lambda\left(1-\frac{1}{\Sxc{\lambda}{c_{1}}}\right)\neq0$.
Hence, the number $\tau$ is uniquely determined by the equality (\ref{eq:8Eq_tau1}).
Then, the vector $\ba$ we calculate by the formula (\ref{eq:vect_a_search_repr}).
And, substituting the delivered values of the parameters $\lambda\in(0,\varepsilon)$,
$\bn\in\BoHa$ and $\ba\in\MHa$ into (\ref{eq:condit(lmbd,n,a)1}),
we obtain the valid equality. Hence, in the case $c_{1}<\infty$,
$c_{2}=\infty$ and $\w_{1}=\oo$, Lemma is proved. 

\textbf{2.} We now turn to the general case, where $\w_{1},\w_{2}$
are arbitrary vectors of the space $\MHa$ such, that $\w_{1}\neq\w_{2}$.
 According to the result, proved in the first item of Lemma, parameters
$\lambda\in(0,\varepsilon)$, $\bn\in\BoHa$ and $\widetilde{\ba}\in\MHa$,
exist such, that $\Www{\lambda,c_{1}}{s,\bn,J}\widetilde{\ba}=\Www{\lambda,c_{2}}{s,\bn,J}\left(\w_{2}-\w_{1}+\widetilde{\ba}\right)$.
Denote, $\ba:=\widetilde{\ba}-\w_{1}$. Then, taking into account,
(\ref{eq:WsnJdef}), we receive the desired equality $\Www{\lambda,c_{1}}{s,\bn,J;\ba}\w_{1}=\Www{\lambda,c_{2}}{s,\bn,J;\ba}\w_{2}$.
       \EndProof  

For $y_{1},y_{2}\in(0,\infty]$, such, that $y_{1},y_{2}\neq\infty$
in the case $y_{1}=y_{2}$ we put: 

\begin{onehalfspace}
\begin{equation}
\sigma\left(y_{1},y_{2}\right)=\begin{cases}
\left(\frac{y_{1}^{-2}+y_{2}^{-2}}{2}\right)^{-\frac{1}{2}}, & y_{1},y_{2}<\infty\\
\sqrt{2}\, y_{1}, & y_{1}<\infty,\: y_{2}=\infty\\
\sqrt{2}\, y_{2}, & y_{1}=\infty,\: y_{2}<\infty
\end{cases}.\label{eq:psi(y1,y2)def}
\end{equation}

\end{onehalfspace}

\BeginLem  \label{Lem:(Wlc1=00003DWlc2)[snj]w}

\begin{onehalfspace}
Suppose, that for some vector $\w\in\MHa$ it holds the equality 
\[
\WsnJ{\lambda,c_{1}}\w=\WsnJ{\lambda,c_{2}}\w,
\]
 where $c_{1},c_{2}\in(0,\infty]$, $\lambda\in(0,\infty]\setminus\left\{ c_{1},c_{2},\sigma\left(c_{1},c_{2}\right)\right\} $,
$s\in\{-1,1\}$, $J\in\UHa$, $\bn\in\BoHa$ with $c_{1}\neq c_{2}$.
~ ~ Then, $\Ttt(\w)=\left\langle \bn,\w\right\rangle =0$. 
\end{onehalfspace}

\EndLem  \vspace{-7mm}

\BeginProof[~ ~\emph{ }Proof of Lemma \em{we divide into a few
cases}] ~

\textbf{Case 1}: $c_{1},c_{2}<\infty$, $\lambda<\infty$. In this
case, by the formula (\ref{eq:WsnJdef}), we get 
\begin{align}
 & \WsnJ{\lambda,c_{1}}\w-\WsnJ{\lambda,c_{2}}\w=\nonumber \\
 & \qquad=\left(\left(\gamma\left(\frac{\lambda}{c_{1}}\right)-\gamma\left(\frac{\lambda}{c_{2}}\right)\right)s\Ttt\left(\w\right)-\left(\frac{\lambda}{c_{1}^{2}}\gamma\left(\frac{\lambda}{c_{1}}\right)-\frac{\lambda}{c_{2}^{2}}\gamma\left(\frac{\lambda}{c_{2}}\right)\right)\left\langle \bn,\w\right\rangle \right)\e+\nonumber \\
 & \qquad\qquad\qquad+\left(\gamma\left(\frac{\lambda}{c_{1}}\right)-\gamma\left(\frac{\lambda}{c_{2}}\right)\right)\left(\lambda\Ttt\left(\w\right)-s\left\langle \bn,\w\right\rangle \right)J\bn,\label{eq:Wlc1[snJ]w-Wlc2[snJ]w}
\end{align}
 where the function $\gamma:[0,\infty)\mapsto\R$ is determined by
the formula (\ref{eq:func_gam_def}).   By conditions of Lemma, $\WsnJ{\lambda,c_{1}}\w-\WsnJ{\lambda,c_{2}}\w=\oo$,
where $\oo$ is zero vector of the space $\MHa$. Hence, the right-hand
side of the equality (\ref{eq:Wlc1[snJ]w-Wlc2[snJ]w}) is equal to
zero vector. Therefore, taking into account orthogonality of the vector
$\e$ to the subspace $\Ha_{1}$ and unitarity of the operator $J$
on the subspace $\Ha_{1}$, we get the following equalities: 
\begin{gather}
\begin{array}{rl}
s\left(\gamma\left(\frac{\lambda}{c_{1}}\right)-\gamma\left(\frac{\lambda}{c_{2}}\right)\right)\Ttt\left(\w\right)-\left(\frac{\lambda}{c_{1}^{2}}\gamma\left(\frac{\lambda}{c_{1}}\right)-\frac{\lambda}{c_{2}^{2}}\gamma\left(\frac{\lambda}{c_{2}}\right)\right)\left\langle \bn,\w\right\rangle  & =0;\\
\left(\gamma\left(\frac{\lambda}{c_{1}}\right)-\gamma\left(\frac{\lambda}{c_{2}}\right)\right)^{\phantom{1}}\left(\lambda\Ttt\left(\w\right)-s\left\langle \bn,\w\right\rangle \right) & =0.
\end{array}\label{eq:LinSys1}
\end{gather}
 According to the conditions of Lemma,  $\lambda>0$ and $\lambda\neq\sigma\left(c_{1},c_{2}\right)=\sqrt{\frac{2}{\frac{1}{c_{1}^{2}}+\frac{1}{c_{2}^{2}}}}$.
Consequently, $\gamma\left(\frac{\lambda}{c_{1}}\right)-\gamma\left(\frac{\lambda}{c_{2}}\right)\neq0$.
Thus, the equalities (\ref{eq:LinSys1}) may be rewritten in the form:
\begin{equation}
\left\{ \begin{array}{rl}
s\left(\gamma\left(\frac{\lambda}{c_{1}}\right)-\gamma\left(\frac{\lambda}{c_{2}}\right)\right)\Ttt\left(\w\right)-\left(\frac{\lambda}{c_{1}^{2}}\gamma\left(\frac{\lambda}{c_{1}}\right)-\frac{\lambda}{c_{2}^{2}}\gamma\left(\frac{\lambda}{c_{2}}\right)\right)\left\langle \bn,\w\right\rangle  & =0;\\
\lambda\Ttt\left(\w\right)-s\left\langle \bn,\w\right\rangle  & =0.
\end{array}\right.\label{eq:LinSys2}
\end{equation}

The system (\ref{eq:LinSys2}) is a system of linear homogeneous equations
relatively the variables $\Ttt\left(\w\right)$ and $\left\langle \bn,\w\right\rangle $.
Determinant of this system is: 
\begin{align*}
\Delta & =-\left[\left(\gamma\left(\frac{\lambda}{c_{1}}\right)-\gamma\left(\frac{\lambda}{c_{2}}\right)\right)-\left(\frac{\lambda^{2}}{c_{1}^{2}}\gamma\left(\frac{\lambda}{c_{1}}\right)-\frac{\lambda^{2}}{c_{2}^{2}}\gamma\left(\frac{\lambda}{c_{2}}\right)\right)\right]=\\
 & =-\left(\boldsymbol{g}\left(\frac{\lambda}{c_{1}}\right)-\boldsymbol{g}\left(\frac{\lambda}{c_{2}}\right)\right),\qquad\text{where}\\
 & \boldsymbol{g}(\xi)=\left(1-\xi^{2}\right)\gamma(\xi)=\sign\left(1-\xi\right)\sqrt{\left|1-\xi^{2}\right|}\qquad(\xi\geq0,\:\xi\neq1).
\end{align*}
 Since the function $\boldsymbol{g}(\xi)=\sign\left(1-\xi\right)\sqrt{\left|1-\xi^{2}\right|}$
is strictly decreasing on $[0,\infty)$, determinant $\Delta$ of
the system (\ref{eq:LinSys2}) is nonzero. Hence, $\Ttt\left(\w\right)=\left\langle \bn,\w\right\rangle =0$,
that was necessary to prove. 

\textbf{Case 2}: $c_{1},c_{2}<\infty$, $\lambda=\infty$. 

In this case, by the formula (\ref{eq:WsnJdef}), we receive:
\begin{align*}
0 & =\WsnJ{\lambda,c_{1}}\w-\WsnJ{\lambda,c_{2}}\w=\\
 & \qquad=-\frac{\left\langle \bn,\w\right\rangle }{c_{1}}\e+c_{1}\Ttt\left(\w\right)J\bn-\left(-\frac{\left\langle \bn,\w\right\rangle }{c_{2}}\e+c_{2}\Ttt\left(\w\right)J\bn\right)=\\
 & \qquad=-\left(\frac{1}{c_{1}}-\frac{1}{c_{2}}\right)\left\langle \bn,\w\right\rangle \e+\left(c_{1}-c_{2}\right)\Ttt\left(\w\right)J\bn.
\end{align*}
And since $c_{1}\neq c_{2}$, taking into account orthogonality of
the vector $\e$ to the subspace $\Ha_{1}$ and unitarity of the operator
$J$ on the subspace $\Ha_{1}$, we get the equality $\Ttt\left(\w\right)=\left\langle \bn,\w\right\rangle =0$. 

\textbf{Case 3}: $c_{1}<\infty$, $c_{2}=\infty$.

By the conditions of Lemma $\lambda\neq c_{2}$. Hence, in this case
we have $\lambda<\infty$. And, according to (\ref{eq:WsnJdef}),
we obtain: 
\begin{align}
0 & =\WsnJ{\lambda,c_{1}}\w-\WsnJ{\lambda,c_{2}}\w=\nonumber \\
 & \qquad=\left(\left(\gamma\left(\frac{\lambda}{c_{1}}\right)-1\right)s\Ttt\left(\w\right)-\frac{\lambda}{c_{1}^{2}}\gamma\left(\frac{\lambda}{c_{1}}\right)\left\langle \bn,\w\right\rangle \right)\e+\nonumber \\
 & \qquad\qquad+\left(\gamma\left(\frac{\lambda}{c_{1}}\right)-1\right)\left(\lambda\Ttt(\w)-s\left\langle \bn,\w\right\rangle \right)J\bn.\label{eq:Wlc1[snJ]w-Wlc2[snJ]w_01}
\end{align}
By the conditions of Lemma, $\lambda>0$ and $\lambda\neq\sigma\left(c_{1},c_{2}\right)=\sqrt{2}\, c_{1}$.
Thus, $\gamma\left(\frac{\lambda}{c_{1}}\right)-1\neq0$. Hence, taking
into account orthogonality of the vector $\e$ to the subspace $\Ha_{1}$
and unitarity of the operator $J$ on the subspace $\Ha_{1}$, from
the equality (\ref{eq:Wlc1[snJ]w-Wlc2[snJ]w_01}) we receive the system
of equations: 
\begin{equation}
\left\{ \begin{array}{rl}
\left(\gamma\left(\frac{\lambda}{c_{1}}\right)-1\right)s\Ttt\left(\w\right)-\frac{\lambda}{c_{1}^{2}}\gamma\left(\frac{\lambda}{c_{1}}\right)\left\langle \bn,\w\right\rangle  & =0\\
\lambda\Ttt(\w)-s\left\langle \bn,\w\right\rangle  & =0.
\end{array}\right.\label{eq:LinSys3}
\end{equation}
The system (\ref{eq:LinSys3}) is a system of linear homogeneous equations
relatively the variables $\Ttt\left(\w\right)$ and $\left\langle \bn,\w\right\rangle $.
Determinant of this system is:
\begin{gather*}
\Delta_{1}=-\left(\left(\gamma\left(\frac{\lambda}{c_{1}}\right)-1\right)-\frac{\lambda^{2}}{c_{1}^{2}}\gamma\left(\frac{\lambda}{c_{1}}\right)\right)=-\left(\boldsymbol{g}\left(\frac{\lambda}{c_{1}}\right)-\boldsymbol{g}\left(0\right)\right).
\end{gather*}
Since, by the conditions of Lemma, $\lambda>0$ and $c_{1}<\infty$,
then $\frac{\lambda}{c_{1}}\neq0$. That is why, $\Delta_{1}\neq0$.
Thus, $\Ttt\left(\w\right)=\left\langle \bn,\w\right\rangle =0$. 

\textbf{Case 4}: $c_{1}=\infty$, $c_{2}<\infty$ is considered similarly
to the case 3. 

\textbf{Case} $c_{1},c_{2}=\infty$ is impossible, because, by the
conditions of Lemma, $c_{1}\neq c_{2}$.    \EndProof 

\BeginNasl \label{Nasl:(Wlc1=00003DWlc2)[snj]w}

Let, $c_{1},c_{2}\in(0,\infty]$, $c_{1}\neq c_{2}$, $s\in\{-1,1\}$,
$J\in\UHa$, $\bn\in\BoHa$. 

Then for any $\w\in\MHa$ and $\varepsilon\in\left(0,\min\left(c_{1},c_{2}\right)\right)$
there exist $\lambda\in(0,\varepsilon)$ and $\ba\in\MHa$, such,
that 
\[
\Www{\lambda,c_{1}}{s,\bn,J;\ba}\w\neq\Www{\lambda,c_{2}}{s,\bn,J;\ba}\w,
\]

\EndNasl 

\BeginProof 

Let us chose any $\ba\in\MHa$ such, that: 
\begin{equation}
\Ttt\left(\w+\ba\right)\neq0.\label{eq:Tau(w-a)neq0}
\end{equation}
Also we chose any number $\lambda\in(0,\varepsilon)\setminus\left\{ \sigma\left(c_{1},c_{2}\right)\right\} $.
If we assume, that $\Www{\lambda,c_{1}}{s,\bn,J;\ba}\w=\Www{\lambda,c_{2}}{s,\bn,J;\ba}\w$,
then, according to (\ref{eq:WsnJdef}), we will obtain: 
\begin{gather*}
\Www{\lambda,c_{1}}{s,\bn,J}(\w+\ba)=\Www{\lambda,c_{2}}{s,\bn,J}(\w+\ba).
\end{gather*}
Hence, by Lemma \ref{Lem:(Wlc1=00003DWlc2)[snj]w}, $\Ttt\left(\w+\ba\right)=0$,
contrary to the correlation (\ref{eq:Tau(w-a)neq0}). Thus, $\Www{\lambda,c_{1}}{s,\bn,J;\ba}\w\neq\Www{\lambda,c_{2}}{s,\bn,J;\ba}\w$.
~     ~ \EndProof 

\BeginProof[Proof of Theorem \ref{Thm:KPT(HBVf)_universPK}]  

\textbf{1}. For any fixed vector $\bn\in\BoHa$ we are going to prove
the equality: 
\begin{equation}
\Www{0;\Vf}{1,\bn,\Ipm{-1,1}{\bn},\oo}=\I_{\MHac},\label{eq:W0Vf=00003DIMHac}
\end{equation}
 where $\I_{\MHac}$ is the the identity operator on $\MHac$, and
\begin{gather*}
\Ipm{\kappa,\mu}{\bn}\, x:=\kappa\Xo{\bn}x+\mu\Xoo{\bn}x,\quad x\in\Ha_{1}\qquad(\bn\in\BoHa,\:\kappa,\mu\in\left\{ -1,1\right\} ).
\end{gather*}
Indeed, according to (\ref{eq:WlVf[snJa]def}), (\ref{eq:WsnJdef}),
for an arbitrary element $\omega=\left(t,\left(x,c\right)\right)\in\MHac$,
we have:
\begin{align*}
 & \hspace{-1cm}\Www{0;\Vf}{1,\bn,\Ipm{-1,1}{\bn},\oo}\omega=\\
 & =\left(\tm{\Www{0,c}{1,\bn,\Ipm{-1,1}{\bn}}\omega^{*}},\:\left(\bs{\Www{0,c}{1,\bn,\Ipm{-1,1}{\bn}}\omega^{*}},c\right)\right)=\\
 & =\left(\tm{\omega^{*}},\:\left(\bs{\omega^{*}},c\right)\right)=\left(t,\:\left(x,c\right)\right)=\omega,
\end{align*}
that was necessary to prove. From the equality (\ref{eq:W0Vf=00003DIMHac})
it follows, that $\I_{\MHac}\in\PTpHC$. Besides this, in accordance
with Remark \ref{Rmk:IdBmmImg}, $\I_{\MHac}[\BBB]=\BBB$. Hence,
by Property \ref{Props:Kim:1}(\ref{enu:PropKim:1:Lk}), we can define
the reference frame: 
\[
\lf_{0}=\left(\I_{\MHac},\I_{\MHac}[\BBB]\right)=\left(\I_{\MHac},\BBB\right)\in\Lk{\KPT{\Ha,\BBB;\Vf}}.
\]

Now, we fix any reference frame $\lf=\left(U,U[\BBB]\right)\in\Lk{\KPT{\Ha,\BBB;\Vf}}$,
where $U=\Www{\lambda,\Vf}{1,\bn,J;\ba}\in\PTpHC$. 

According to Properties \ref{Props:Kim:1}(\ref{enu:PropKim:1:FrameComponents},~\ref{enu:PropKim:1:Unif!}),
we obtain: 
\begin{align}
\Mk(\lf) & =\R\times\Zk\left(\Haa\right)=\R\times\Ha=\MHa;\label{eq:Mk(l)}\\
\unn{\lf_{0}}{\lf}{\omega} & =U\left(\arc{\I\,_{\MHac}}\omega\right)=U\omega=\Www{\lambda,\Vf}{1,\bn,J;\ba}\omega\label{eq:KPTHC!<l0_to_l>1}\\
 & \qquad\qquad(\forall\omega\in\BS\left(\lf_{0}\right)=\BSB\subseteq\MHac.\nonumber 
\end{align}

Using Property \ref{Props:Kim:1}(\ref{enu:PropKim:1:FrameComponents})
as well as equality (\ref{eq:def:PTmprj+q}), for an elementary-time
state~ $\omega=\left(t,\left(x,c\right)\right)\in\BS(\lf)$ we get:
\begin{align}
\pk{\lf}(\omega) & =\left(\tm{\omega},\mathbf{q}(\bs{\omega})\right)=\left(t,\mathbf{q}\left(\left(x,c\right)\right)\right)=\left(t,x\right)=\omega^{*}.\label{eq:KPTHC!<l0_to_l>2}
\end{align}
 Hence, using Definition \ref{Def:PeretvKoord} (item \ref{enu:RealPeretvKoord})
and equality (\ref{eq:Wlc[snJa]w*}), we deduce: 
\begin{align}
\PKK{\lf_{0}}{\lf}(\omega) & =\pk{\lf}\left(\unn{\lf_{0}}{\lf}{\omega}\right)=\left(\Www{\lambda,\Vf}{1,\bn,J;\ba}\omega\right)^{*}=\nonumber \\
 & \qquad=\Www{\lambda,c}{1,\bn,J;\ba}\omega^{*}\quad\left(\forall\omega\in\BS\left(\lf_{0}\right)=\BSB\subseteq\MHac\right).\label{eq:KPTHC!<l0_to_l>3}
\end{align}

\textbf{2}. By conditions of Theorem a number $\eta>0$ exists such,
that $\Vf\subseteq[\eta,\infty)$. 

\textbf{2.1}. Suppose, that there exist elementary states $\xxv_{1}=\left(x_{1},c_{1}\right),\:\xxv_{2}=\left(x_{2},c_{2}\right)\in\BsB$
such, that $c_{1}\neq c_{2}$. Since, by Property \ref{Props:BMM}(\ref{enu:BMMProp(Bs)}),
$\BsB=\left\{ \bs{\omega}\,|\,\omega\in\BSB\right\} $, then there
exist elementary-time states of kind $\omega_{1}=\left(t_{1},\xxv_{1}\right)=\left(t_{1},\left(x_{1},c_{1}\right)\right)\in\BSB$,
$\omega_{2}=\left(t_{2},\xxv_{2}\right)=\left(t_{2},\left(x_{2},c_{2}\right)\right)\in\BSB$.
 Now, we consider two cases.  

\textbf{Case 2.1.1}: $\omega_{1}^{*}\neq\omega_{2}^{*}$. Consider
any fixed operator $J_{1}\in\UHa$. By Lemma \ref{Lem:(Wlc1[snj]w1=00003DWlc2[snj]w2)},
there exist $\lambda_{1}\in(0,\eta)$, $\bn_{1}\in\BoHa$ and $\ba_{1}\in\MHa$,
such, that 
\begin{equation}
\Www{\lambda_{1},c_{1}}{1,\bn_{1},J_{1};\ba_{1}}\omega_{1}^{*}=\Www{\lambda_{1},c_{2}}{1,\bn_{1},J_{1};\ba_{1}}\omega_{2}^{*}.\label{eq:Wlc1[snj]w1=00003DWlc2[snj]w2}
\end{equation}
Let us introduce the reference frame: 
\begin{align*}
\lf_{1} & :=\left(U_{1},U_{1}[\BBB]\right)\in\Lk{\KPT{\Ha,\BBB;\Vf}},\quad\text{where}\\
U_{1} & :=\Www{\lambda_{1};\Vf}{1,\bn_{1},J_{1},\ba_{1}}\in\PTpHC.
\end{align*}
 According to (\ref{eq:KPTHC!<l0_to_l>3}), and (\ref{eq:Wlc1[snj]w1=00003DWlc2[snj]w2}),
we receive: 
\[
\PKK{\lf_{0}}{\lf_{1}}\left(\omega_{1}\right)=\Www{\lambda_{1},c_{1}}{1,\bn_{1},J_{1};\ba_{1}}\omega_{1}^{*}=\Www{\lambda_{1},c_{2}}{1,\bn_{1},J_{1};\ba_{1}}\omega_{2}^{*}=\PKK{\lf_{0}}{\lf_{1}}\left(\omega_{2}\right).
\]
 From the other hand, by the formula (\ref{eq:KPTHC!<l0_to_l>2}),
we obtain $\pk{\lf_{0}}\left(\omega_{1}\right)=\omega_{1}^{*}\neq\omega_{2}^{*}=\pk{\lf_{0}}\left(\omega_{2}\right)$.
Thus, for the elementary-time states $\omega_{1}$,$\omega_{2}$ we
have $\PKK{\lf_{0}}{\lf}\left(\omega_{1}\right)=\PKK{\lf_{0}}{\lf}\left(\omega_{2}\right)$,
while $\pk{\lf_{0}}\left(\omega_{1}\right)\neq\pk{\lf_{0}}\left(\omega_{2}\right)$.
Hence, by Theorem \ref{Thm:GPKexistenceCriteria}, the reference frames
$\lf_{0}$ and $\lf$ do not allow universal coordinate transform.
Therefore, in accordance with Assertion \ref{As:GPKkinSpaceEquivalence},
item 2, the kinematic set $\KPT{\Ha,\BBB;\Vf}$ do not allow universal
coordinate transform in this case. 

\textbf{Case 2.1.2}: $\omega_{1}^{*}=\omega_{2}^{*}$. Consider any
fixed operator $J_{2}\in\UHa$ and vector $\bn_{2}\in\BoHa$. According
to Corollary \ref{Nasl:(Wlc1=00003DWlc2)[snj]w}, there exist $\lambda_{2}\in(0,\eta)$
and $\ba_{2}\in\MHa$, such, that 
\begin{equation}
\Www{\lambda_{2},c_{1}}{1,\bn_{2},J_{2};\ba_{2}}\omega_{1}^{*}\neq\Www{\lambda_{2},c_{2}}{1,\bn_{2},J_{2};\ba_{2}}\omega_{2}^{*}.\label{eq:Wlc1[snj]w1<>Wlc2[snj]w2}
\end{equation}
Let us consider the reference frame: 
\begin{align*}
\lf_{2}: & =\left(U_{2},U_{2}[\BBB]\right)\in\Lk{\KPT{\Ha,\BBB;\Vf}},\quad\text{where}\\
U_{2} & =\Www{\lambda_{2};\Vf}{1,\bn_{2},J_{2},\ba_{2}}\in\PTpHC.
\end{align*}
 According to (\ref{eq:KPTHC!<l0_to_l>3}), and (\ref{eq:Wlc1[snj]w1<>Wlc2[snj]w2}),
we receive: 
\[
\PKK{\lf_{0}}{\lf_{2}}\left(\omega_{1}\right)=\Www{\lambda_{2},c_{1}}{1,\bn_{2},J_{2};\ba_{2}}\omega_{1}^{*}\neq\Www{\lambda_{2},c_{2}}{1,\bn_{2},J_{2};\ba_{2}}\omega_{2}^{*}=\PKK{\lf_{0}}{\lf_{2}}\left(\omega_{2}\right).
\]
 From the other hand, by the formula (\ref{eq:KPTHC!<l0_to_l>2}),
we obtain: $\pk{\lf_{0}}\left(\omega_{1}\right)=\omega_{1}^{*}=\omega_{2}^{*}=\pk{\lf_{0}}\left(\omega_{2}\right)$.

Thus, for the elementary-time states $\omega_{1}$,$\omega_{2}$ we
have $\PKK{\lf_{0}}{\lf_{2}}\left(\omega_{1}\right)\neq\PKK{\lf_{0}}{\lf_{2}}\left(\omega_{2}\right)$,
while $\pk{\lf_{0}}\left(\omega_{1}\right)=\pk{\lf_{0}}\left(\omega_{2}\right)$.
Hence, by Theorem \ref{Thm:GPKexistenceCriteria}, the reference frames
$\lf_{0}$ and $\lf_{2}$ do not allow universal coordinate transform.
Therefore, in accordance with Assertion \ref{As:GPKkinSpaceEquivalence},
item 2, the kinematic set $\KPT{\Ha,\BBB;\Vf}$ does not allow universal
coordinate transform. 

\textbf{\emph{Thus}}, if the kinematic set $\KPT{\Ha,\BBB;\Vf}$ allows
universal coordinate transform, then there not exist elementary states
$\xxv_{1}=\left(x_{1},c_{1}\right),\:\xxv_{2}=\left(x_{2},c_{2}\right)\in\BsB$
such, that $c_{1}\neq c_{2}$. 

\textbf{2.2}. Now we suppose, that in base changeable set $\BBB$
there not exist elementary states $\xxv_{1}=\left(x_{1},c_{1}\right),\:\xxv_{2}=\left(x_{2},c_{2}\right)\in\BsB$
such, that $c_{1}\neq c_{2}$. Under this assumption a number $c_{0}\in\Vf$
must exist such, that arbitrary elementary state $\xxv\in\BsB$ can
be represented in the form: $\xxv=\left(x,c_{0}\right)$, where $x\in\Ha$.
Chose any reference frame: 
\begin{align*}
\lf: & =\left(U,U[\BBB]\right)\in\Lk{\KPT{\Ha,\BBB;\Vf}},\quad\text{where}\\
U & =\Www{\lambda;\Vf}{1,\bn,J,\ba}\in\PTpHC.
\end{align*}
According to (\ref{eq:KPTHC!<l0_to_l>3}), (\ref{eq:KPTHC!<l0_to_l>2}),
for arbitrary elementary-time state $\omega=\left(t,\left(x,c_{0}\right)\right)\in\BS\left(\lf_{0}\right)=\BSB$
we obtain: 
\begin{align*}
\PKK{\lf_{0}}{\lf}\left(\omega\right) & =\Www{\lambda,c_{0}}{1,\bn,J;\ba}\omega^{*}=\Www{\lambda,c_{0}}{1,\bn,J;\ba}\left(\pk{\lf_{0}}(\omega)\right),
\end{align*}
where $\Www{\lambda,c_{0}}{1,\bn,J;\ba}$ is a bijection from $\MHa$
onto $\MHa$ (and, by ($\ref{eq:Mk(l)}$), $\Www{\lambda,c_{0}}{1,\bn,J;\ba}$
is a bijection from $\Mk\left(\lf_{0}\right)$ onto $\Mk(\lf)$).
Hence, in accordance with Definition \ref{Def:PeretvKoord}, the mapping
$\Www{\lambda,c_{0}}{1,\bn,J;\ba}$ is universal coordinate transform
from $\lf_{0}$ to $\lf$. Consequently, the reference frames $\lf_{0}$
and $\lf$ allow universal coordinate transform, ie $\lf_{0}\gteq\lf$
(for any reference frame $\lf\in\Lk{\KPT{\Ha,\BBB;\Vf}}$).  Thus,
by Assertion \ref{As:GPKkinSpaceEquivalence}, kinematic set $\KPT{\Ha,\BBB;\Vf}$
allows universal coordinate transform.     \EndProof 

Similarly to Theorem \ref{Thm:KPT(HBVf)_universPK} it can be proved
the following theorem. 

\BeginThm \label{Thm:KPT0(HBVf)_universPK} 

Let the set of forbidden velocities $\Vf\subseteq(0,\infty]$ be separated
from zero (ie there exists a number $\eta>0$ such, that $\Vf\subseteq[\eta,\infty]$). 

Kinematic set $\KPTn{\Ha,\BBB;\Vf}$ allows universal coordinate transform
if and only if there don't exist elementary states $\xxv_{1}=\left(x_{1},c_{1}\right),\:\xxv_{2}=\left(x_{2},c_{2}\right)\in\BsB$
such, that $c_{1}\neq c_{2}$. 

\EndThm 

Note, that theorems \ref{Thm:KPT(HBVf)_universPK} and \ref{Thm:KPT0(HBVf)_universPK}
were announced in the paper \cite{MyTmm09(DAN)}.

\section{Conclusions. }

Development of kinematic theories of tachyon movement (which is especially
intensified in the recent years) generates the problem of building
a new mathematical apparatus, which would allow to investigate evolution
of physical systems in a framework of different laws of kinematics.
Concerning to the given problem in this paper the following results
are obtained: 
\begin{enumerate}
\item The definitions of actual and universal coordinate transform in kinematic
sets are presented. 
\item The kinematic sets of kind $\KP{\Ha,\BBB,c}$, $\KPn{\Ha,\BBB,c}$,
$\KPT{\Ha,\BBB,c}$, $\KPTn{\Ha,\BBB,c}$ are constructed. These kinematic
sets represent mathematically strict models of evolution of physical
systems in the framework of kinematics of special relativity theory
as well as it's tachyon extension based on the generalized Lorentz-Poincare
transformations (in the sense of E. Recami). 
\item The kinematic sets of kind $\KPT{\Ha,\BBB;\Vf}$ and $\KPTn{\Ha,\BBB;\Vf}$
are constructed. These kinematic sets may simulate evolution of physical
systems under the condition of hypothesis on existence of particle-dependent
velocity of light. 
\item It is proved, that the kinematic sets of type $\KP{\Ha,\BBB,c}$,
$\KPn{\Ha,\BBB,c}$, $\KPT{\Ha,\BBB,c}$, $\KPTn{\Ha,\BBB,c}$ allow
universal coordinate, whereas in the kinematic sets $\KPT{\Ha,\BBB;\Vf}$,
$\KPTn{\Ha,\BBB;\Vf}$ universal coordinate transform can not exist
in non-trivial cases. 
\end{enumerate}


\begin{thebibliography}{10}
\bibitem{OPERA01}T. Adam et al., {[}OPERA Collaboration{]}. \emph{Measurement
of the neutrino velocity with the OPERA detector in the CNGS beam}.
Preprint: arXiv:1109.4897v2 {[}hep-ex{]}, (Sep 2011). 

\bibitem{Bilaniuk01}O.-M. P. Bilaniuk, V. K. Deshpande, E. C. G.
Sudarshan. \emph{``Meta'' Relativity}. American Journal of Physics.
\textbf{30} (10), (1962), 718.

\bibitem{Bilaniuk02} O.-M. P. Bilaniuk, E. C. G. Sudarshan. \emph{Particles
beyond the Light Barrier}. Physics Today. \textbf{22}, (5), (1969),
43–51.

\bibitem{BaccettiTateVisser}Valentina Baccetti, Kyle Tate, Matt Visser.
\emph{Inertial frames without the relativity principle}. Journal of
High Energy Physics. \textbf{2012}, (5), (2012). 

\bibitem{Recami1}E. Recami. \emph{Classical Tachyons and Possible
Applications}.~ Riv. Nuovo Cim. \textbf{9}, (s.~3, N~6), (1986),
1-178. 

\bibitem{Ricardo1}Ricardo S. Vieira. \emph{An Introduction to the
Theory of Tachyons.} Preprint: arXiv:1112.4187v2, (2012). 

\bibitem{Hill_Cox}James M. Hill, Barry J. Cox. \emph{Einstein's special
relativity beyond the speed of light}.~ Proc. of the Royal Society.
\textbf{468}, (2148), (December 2012), 4174-4192. 

\bibitem{MyTmmTaxion01}Grushka Ya.I. \emph{Tachyon Generalization
for Lorentz Transforms}.~ Methods of Functional Analysis and Topology.
\textbf{20}, (2), (2013), P. 127-145.

\bibitem{MyTmmTaxion02}Grushka Ya.I. \emph{Algebraic Properties of
Tachyon Lorentz Transforms}~//~Proceedings of Institute of Mathematics
NAS of Ukraine. \textbf{10} (2), (2013), 138-169, (In Ukrainian, English
translation is available at {\small http:// www.researchgate.net/publication/257933423\_Algebraic\_Properties\_of\_Tachyon\_Lorentz\_Transforms}). 

\bibitem{MyTmm01}Grushka Ya.I.  \emph{Changeable sets and their properties}.
Reports of the National Academy of Sciences of Ukraine. \textbf{2012},
(5), (2012), 12-18, (In Ukrainian,  {\small http://www.researchgate.net/publication/236120448\_Changeable\_sets\_and\_their\_properties}). 

\bibitem{MyTmm03}Grushka Ya.I. \emph{Abstract concept of changeable
set.} Preprint arXiv:1207.3751v1, (2012).

\bibitem{MyTmm04}Grushka Ya.I.\emph{ Visibility in changeable sets.
}Proceedings of Institute of Mathematics NAS of Ukraine. \textbf{9},
(2), (2012), 122-145, (In Ukrainian, {\small http://www.researchgate.net/publication/236217050\_Visibility\_in\_changeable\_sets}). 

\bibitem{MyTmm06}Grushka Ya.I.\emph{ Changeable sets and their application
for the construction of tachyon kinematics}. Proceedings of Institute
of Mathematics NAS of Ukraine. \textbf{11}, (1), (2014), 192-227,
(In Ukrainian). 

\bibitem{MyTmm05(YMJ)}Grushka Ya.I. \emph{Base changeable sets and
mathematical simulation of the evolution of systems}. Ukrainian Mathematical
Journal. \textbf{65}, (9), (2014), 1332-1353. 

\bibitem{Birkhoff}Birkhoff, Garrett. \emph{Lattice theory}. New York,
(1967). 

\bibitem{MyTmm07}Grushka Ya.I. \emph{Evolutionary expansion and analogs
of the union operation for base changeable sets.} Proceedings of Institute
of Mathematics NAS of Ukraine. \textbf{11}, (2), (2014), (In Ukrainian). 

\bibitem{MyTmm08}Grushka Ya.I. \emph{Existence criteria for universal
coordinate transforms in kinematic changeable sets}. Bukovinian Mathematical
Journal. \textbf{2}, (2-3), (2014), 59-71, (In Ukrainian, English
translation is available at {\small http://www.researchgate.net/publication/}\\
{\small $\,$ 270647695\_Existence\_Criteria\_for\_Universal\_Coordinate\_Transforms\_in\_Kinematic\_Changeable\_Sets}). 

\bibitem{Naimark1}M. A. Naimark, \emph{Linear Representations of
the Lorentz Group, }International Series of Monographs in Pure and
Applied Mathematics, Vol. \textbf{63}, (XIV), Oxford/London/Edinburgh/New
York/Paris/Frankfurt, (1964).

\bibitem{Myoler01} C. M$\ddot{\textrm{o}}$ller. \emph{The Theory
of Relativity}. (Oxford), 1972. 

\bibitem{BaccettiTateVisser02}Valentina Baccetti, Kyle Tate, and
Matt Visser. \emph{Lorentz violating kinematics: Threshold theorems}.
Journal of High Energy Physics. \textbf{2012}, (3), (March 2012). 

\bibitem{BaccettiTateVisser03}Valentina Baccetti, Kyle Tate, and
Matt Visser. \emph{Inertial frames without the relativity principle:
breaking Lorentz symmetry}. Preprint: arXiv:1302.5989, (2013).

\bibitem{EoloDiCasola01}Eolo Di Casola. \emph{Sieving the Landscape
of Gravity Theories}. Thesis submitted for the degree of Ph.D. SISSA,
(2014).

\bibitem{S_Liberati01}S. Liberati. \emph{Tests of Lorentz invariance:
a 2013 update}. Class. Quant. Grav. \textbf{30}, (13), (2013), 133001
(arXiv:~1304.5795 {[}gr-qc{]}).

\bibitem{Gao_shan01}Gao shan. \emph{How to realize quantum superluminal
communication}. Preprint: arXiv:quant-ph/9906116v2, (8 Jul 1999). 

\bibitem{Kholmetskii01}A. L. Kholmetskii, Oleg V. Missevitch, Roman
Smirnov Rueda, T. Yarman. \emph{The special relativity principle and
superluminal velocities}. Physics essays. \textbf{25}, (4), (2012),
621-626.

\bibitem{Glashow01}S. R. Coleman, S. L. Glashow. \emph{Cosmic ray
and neutrino tests of special relativity}. Physics Letters B, \textbf{405},
(1997), 249-252. (arXiv:~hep-ph/9703240). 

\bibitem{Glashow02}S. R. Coleman, S. L. Glashow. \emph{High-energy
tests of Lorentz invariance}. Physical Review D. \textbf{59}, (116008),
(1999), 249-252. (arXiv:~hep-ph/9703240). 

\bibitem{Drago01}Alessandro Drago, Isabella Masina, Giuseppe Pagliara,
and Raffaele Tripiccione. \emph{The Hypothesis of Superluminal Neutrinos:
comparing OPERA with other Data}. Europhysics Letters. \textbf{97},
(2), (2012) (arXiv:1109.5917 {[}hep-ph{]}).

\bibitem{Gertov01}Helene Gertov. \emph{Lorentz Violations.} Published
by University of Southern Denmark. (August, 2012). 

\bibitem{Kazarian01}G. Ter-Kazarian. \emph{Extended Lorentz code
of a superluminal particle}. Preprint: arXiv:1202.0469 {[}physics.gen-ph{]},
(2012). 

\bibitem{MyTmm09(DAN)}Grushka Ya.I. \emph{Coordinate Transforms in
Kinematic Changeable Sets}. Reports of the National Academy of Sciences
of Ukraine. (03), (2015), 24-31. (In Ukrainian, English translation
is available at {\small http://www.researchgate.net/publication/}\\
{\small $\,$ 274374774\_Coordinate\_Transforms\_in\_Kinematic\_Changeable\_Sets}). 

\end{thebibliography}
\end{document}